\documentclass[twoside]{IEEEtran}

\usepackage{epsf}
\usepackage{graphicx}
\DeclareGraphicsExtensions{.png}
\usepackage{caption}
\usepackage{subcaption}
\usepackage{xcolor}
\usepackage{amsmath}
\usepackage{amsfonts}
\usepackage{amssymb}
\usepackage{cuted}

\newcommand{\ul}[1]{\underline{#1}}

\newcommand{\dul}[1]{\underline{\underline{#1}}}

\newcommand{\eps}{\epsilon}

\newcommand{\sinc}{{\rm sinc}}

\newcommand{\bea}{\begin{eqnarray}}

\newcommand{\eea}{\end{eqnarray}}

\newcommand{\rmB}{{\rm B}}

\newcommand{\rmc}{{\rm c}}

\newcommand{\rmd}{{\rm d}}

\newcommand{\rmD}{{\rm D}}

\newcommand{\rmE}{{\rm E}}

\newcommand{\rmj}{{\rm j}}

\newcommand{\rmH}{{\rm H}}

\newcommand{\rmIm}{{\rm Im}}

\newcommand{\rmM}{{\rm M}}

\newcommand{\rmP}{{\rm P}}

\newcommand{\rmopt}{{\rm opt}}

\newcommand{\rmS}{{\rm S}}

\newcommand{\rmT}{{\rm T}}

\newcommand{\rmU}{{\rm U}}

\newcommand{\cE}{{\cal E}}

\newcommand{\cH}{{\cal H}}

\newcommand{\cD}{{\cal D}}

\newcommand{\cB}{{\cal B}}

\newcommand{\cJ}{{\cal J}}

\newcommand{\cF}{{\cal F}}

\newcommand{\cP}{{\cal P}}

\newcommand{\cT}{{\cal T}}

\newcommand{\dee}{{\rm d}}

\begin{document}

%\hyphenate{instan-ta-neous}

%TITLEPAGE

\title{Average Linear and Angular Momentum 
and Power of Random Fields Near a Perfectly 
\\
Conducting Boundary
}

\author{
{Luk R. Arnaut
%, 
%\it Senior Member IEEE \rm 
and Gabriele Gradoni
%, \it Member IEEE \rm
}
}

%\address{Laboratory of Stochastic Electromagnetism and Wave Complexity, School of Electronic Engineering and Computer Science, Queen Mary University, London E1 4FZ, United Kingdom, e-mail: l.arnaut@qmul.ac.uk\\
%School of Mathematical Sciences, University of Nottingham, Nottingham, U.K., e-mail: gabriele.gradoni@nottingham.ac.uk}

\maketitle

%ABSTRACT

\begin{abstract}
The effect of a perfectly conducting planar boundary on the average linear momentum (LM), angular (momentum (AM), and their power for a time-harmonic statistically isotropic random field is analyzed. 
These averages are purely imaginary and their magnitude decreases in a damped oscillatory manner with distance from the boundary. At discrete quasi-periodic distances and frequencies, the average LM and AM attain their free-space value. Implications for the optimal placement or tuning of power and field sensors are analyzed. 
Conservation of the flux of the mean LM and AM
with respect to the difference of the average electric and magnetic energies and the radiation stresses via the Maxwell stress dyadic is demonstrated. 
The second-order spatial derivatives of differential radiation stress can be directly linked to the electromagnetic energy imbalance.
Analytical results are supported by Monte Carlo simulation results.
As an application, performance based estimates for the working volume of a reverberation chamber are obtained.
In the context of multiphysics compatibility, mechanical self-stirred reverberation is proposed as an exploitation of electromagnetic stress.
\end{abstract}

\section{Introduction\label{sec:intro}}
Angular momentum (AM) in electromagnetics was first theorized by Poynting \cite{poyn1909} and consists of spin (or intrinsic) and orbital contributions. 
Spin angular momentum (SAM) is represented by familiar left- and right-hand circular field polarizations, as demonstrated experimentally by Beth \cite{beth1936}. 
Recently, it was found that light waves with a helicoidal phase front exhibit another source of AM whose magnitude may be several times larger than that of SAM \cite{alle1992}. 
This additional part was named orbital angular momentum (OAM) and depends on the transverse spatial field variation.
OAM can be identified with a phase front rotating helicoidally around the axial direction of energy propagation $\ul{k}$ that governs linear momentum (LM).

More recently, OAM has gained renewed interest in radio engineering because of its potential for increasing the number of degrees of freedom and channel capacity of MIMO wireless communications systems \cite{thide2015}. Its mechanism is the generation of orthogonal transmission modes with several different azimuthal phase patterns (integer multiples of $2\pi$) and associated AM power flux, while maintaining the overall point-to-point LM power flux (Poynting vector). 

In the OAM literature, the focus has been almost universally on unbounded propagation in free space. Real scenarios, however, involve antennas near a ground plane, scatterers, multipath reflection, etc.
This raises the question to what extent LM and (O)AM can then be preserved or detected.
One motivation for studying OAM near a boundary is the use of reverberation chambers  for evaluating (O)AM properties. 
Furthermore, OAM may arise unintentionally, e.g., from a particular phase distribution of edge currents around apertures or circuit loops. This may give rise to azimuthal AM power in the near field in a manner that may not be detectable by wire or loop sensors for conventional LM power.

LM and AM may also be instrumental in the conversion between electromagnetic (EM) energy and other forms of physical energy, in what may be termed {\it multiphysics compatibility} (MPC). In particular, MPC includes coupling phenomena between EM fields and their induced or external forces on material bodies via radiation stress and shear, i.e., electromagnetic-mechanical compatibility (EMMC); cf. Sec. \ref{sec:MST}. 
Another subdomain of MPC is electromagnetic-thermodynamic compatibility (EMTC), originating from ohmic losses that are already incorporated in EM constitutive relations. 
EMMC or EMTC may be significant in e.g. vacuum chambers, aerospace, applications involving significant EM forces on small or light objects with large surface-to-volume ratios (``smart dust''), nanoscale components including VLSI circuits \cite{slep2015}, etc., 
and in high-power applications where mechanical effects of EM forces and heating may lead to deformation and arcing in extreme cases.
A basic tenet of MPC is that excitation in one subdomain (e.g., EM) should preserve functionality and nominal conditions of operation that remain within tolerance levels also for other MPC subdomains. 
Here too reverberation chambers offer a primary test bed, because of their ability to generate high field strengths and power densities, inducing extreme mechanical and thermal MPC effects.

In this paper, we analyze the effect of a rigid infinite planar perfect electrically conducting (PEC) surface on the LM and AM power of an ideal isotropic random field incident from a half-space. The analysis extends earlier results for LM and AM in unbounded free space \cite{paololegacy} and for energy density near a PEC plane \cite{dunn1990}, \cite{arnaTEMCmay2006}.
The approach differs from traditional studies of OAM, in which a helicoidal wavefront is considered from the start.
Instead, we analyze to what extent statistical AM may be induced or reduced as a result of interaction with a PEC boundary. 
A previously developed methodology  \cite{arnaPREmar2006} based on local angular spectral plane-wave expansions near an impedance boundary is applied and extended. 
Throughout this paper, general time-dependent EM quantities are shown in roman type; time-harmonic quantities have a suppressed $\exp(\rmj \omega t)$ dependence and are denoted in italics.

\section{Linear and Angular Momentum,\\ Power and Energy}
To establish notions and notations, the definitions of LM and AM for general time-dependent and harmonic fields, and their connection to EM energy and power are briefly reviewed.

For spatiotemporal fields $\ul{\rmE}(\ul{r},t)$ and $\ul{\rmB}(\ul{r},t)$, the 
(real) 
local instantaneous AM density with reference to a location $\ul{r}_0$ is \cite[ch. 6]{jack1975}, \cite[ch. 1]{kong1990}
\bea
\ul{\rmM} (\ul{r}, \ul{r}_0, t) \stackrel{\Delta}{=} (\ul{r} - \ul{r}_0) \times \ul{\rmP} = \mu_0 \eps_0 (\ul{r} - \ul{r}_0) \times \ul{\rmS}
\label{eq:Mgeneral_def}
\eea
where 
\bea
\ul{\rmP} (\ul{r}, t) \stackrel{\Delta}{=} \ul{\rmD} (\ul{r}, t) \times \ul{\rmB} (\ul{r}, t) = \ul{\rmS} (\ul{r}, t) / {\rmc}^2
\eea
is the LM density and $\ul{\rmS}(\ul{r}, t) = \ul{\rmE} (\ul{r}, t) \times \ul{\rmH} (\ul{r}, t)$ is the local LM power flux density (Poynting vector).
With $\ul{\nabla} = (\partial/\partial r)\ul{1}_r$,
%Faraday's equation 
$\ul{\nabla} \times \ul{\rmE} = - \partial\ul{\rmB} / \partial t$, and 
%Gauss's law 
$\ul{\nabla} \cdot \ul{\rmD} = \rho$, (\ref{eq:Mgeneral_def}) becomes
\bea
{\ul{\rmM} (\ul{r}, \ul{r}_0,t)} &=& \eps_0 (\ul{r} - \ul{r}_0) \times \int^t \left [ \ul{\nabla} (\ul{\rmE}\cdot \ul{\rmE}) - (\ul{\nabla} \cdot \ul{\rmE}) \ul{\rmE} \right ] \dee t \nonumber\\
&=& (\ul{r} - \ul{r}_0) \times \int^t \left ( \ul{\nabla} {\rmU}_{em} - \rho \ul{\rmE} \right ) {\rmd}t
\label{eq:Mgeneral}
\eea
where ${\rmU}_{em}(\ul{r}, t) = {\rmU}_{e} + {\rmU}_{m} \stackrel{\Delta}{=} \ul{\rmE} \cdot \ul{\rmD} / 2 + \ul{\rmB} \cdot \ul{\rmH} / 2 $.
The electric and magnetic energy densities ${\rmU}_e$ and ${\rmU}_m$ are quadratic functions of the EM field that are of purely electric and magnetic type, whereas $\ul{\rmS}$ is of mixed types.

For time-harmonic fields $\ul{E}(\ul{r},\omega)$ and $\ul{B}(\ul{r},\omega)$, the complex Poynting vector is
$\ul{S}(\ul{r},\omega) \stackrel{\Delta}{=} (\ul{E} \times \ul{H}^*) / 2$, with Re$(\ul{S}) = \overline{\ul{\rmE} \times \ul{\rmH}}$ representing the time averaged LM power flux density.  
The corresponding (complex) AM density is 
\bea
\ul{M} (\ul{r}, \ul{r}_0, \omega) 
       &\stackrel{\Delta}{=}& (\ul{r} - \ul{r}_0) \times \ul{P} = \mu_0 \eps_0 (\ul{r} - \ul{r}_0) \times \ul{S}\nonumber\\
       &=& (\ul{r} - \ul{r}_0) \times \frac{\rmj\eps_0}{2\omega} \left [ \ul{\nabla} |\ul{E}|^2 - (\ul{\nabla} \cdot \ul{E}) \ul{E}^* \right ]
.~~
\label{eq:Mcomplex}
\eea
For a plane wave, (\ref{eq:Mgeneral}) with $\ul{\nabla} = - \rmj \ul{k} \equiv - \rmj (\omega/\rmc) \ul{1}_k$ yields
\bea
\ul{M} (\ul{r}, \ul{r}_0, \ul{k})
       = (\ul{r} - \ul{r}_0) \times \left ( \frac{U_{em}}{\rmc} \ul{1}_k - \rmj \frac{\rho}{2\omega} \ul{E}^* \right )
       \label{eq:Mharmonic}
\eea
where $U_{em}(\ul{r},\ul{k}) = U_e + U_m = \eps_0 \ul{E} \cdot \ul{E}^* / 4 + \mu_0 \ul{H} \cdot \ul{H}^* / 4$ \cite{kong1990}.
It is further assumed that $\ul{r}_0 = \ul{0}$ and $\rho \ll k||\ul{D}||$, so that the second term in (\ref{eq:Mharmonic}) can be neglected.
In Cartesian coordinates, 
\bea
&~&\hspace{-0.5cm} \ul{P} (\ul{r},\omega) = \frac{\mu_0\eps_0}{2} \left [ (E_y H^*_z - E_z H^*_y) \ul{1}_x + (E_z H^*_x \right. \nonumber\\
&~&~~~~~~~~ \left. 
- E_x H^*_z) \ul{1}_y 
+ (E_x H^*_y - E_y H^*_x) \ul{1}_z \right ]
\label{eq:avgS}\\
%\eea
%\bea
&~& \hspace{-0.5cm}\ul{M} (\ul{r},\ul{0},\omega) = \frac{\mu_0\eps_0}{2} \times \nonumber\\
&~& \left \{ \left [ y (E_x H^*_y - E_y H^*_x) - z (E_z H^*_x - E_x H^*_z) \right ] \ul{1}_x \right . \nonumber\\ 
&~&  \left. + \left [ z (E_y H^*_z - E_z H^*_y) - x (E_x H^*_y - E_y H^*_x) \right ] \ul{1}_y \right . \nonumber\\ 
&~&  \left. + \left [ x (E_z H^*_x - E_x H^*_z) - y (E_y H^*_z - E_z H^*_y) \right ] \ul{1}_z \right \}.~~
\label{eq:avgM}
\eea

The total AM can be decomposed into SAM and OAM contributions \cite[ch. 7]{jack1975} in dual-symmetrized form \cite{berr2009} as 
\bea
\ul{M}(\ul{r},\ul{r}_0,\omega) &=& \frac{1}{4}
\left [ \eps_0 ( \ul{E} \times \ul{A}^* ) + \mu_0 ( \ul{H} \times \ul{F}^* ) \right ] \nonumber\\
&~& + (\ul{r}-\ul{r}_0) \times
\frac{1}{2} \left [ 
\eps_0 (\ul{\nabla} \, \ul{A}) \cdot \ul{E}^* 
\right. \nonumber\\
&~& \left . +
\mu_0 (\ul{\nabla} \, \ul{F}) \cdot \ul{H}^* 
\right ]
\nonumber\\
&\stackrel{\Delta}{=}& \ul{M}_s (\ul{r}, \omega) + \ul{M}_o (\ul{r},\ul{r}_0,\omega)
\label{eq:Ms_plus_Mo}
\eea
where $(\ul{\nabla} \, \ul{A}) \cdot \ul{E}^* \equiv \ul{\nabla} (\ul{A} \cdot \ul{E}^*)$, etc. This identification requires knowledge of the magnetic and electric vector potentials $\ul{A}(\ul{r}|\ul{r}^\prime)$ and $\ul{F}(\ul{r}|\ul{r}^\prime)$, and hence the spatial distributions of electric and magnetic source currents $\ul{J}(\ul{r}^\prime)$ and $\ul{K}(\ul{r}^\prime)$ need to be specified, respectively.
In source-free regions, $\ul{A}$ and $\ul{F}$ in (\ref{eq:Ms_plus_Mo}) are replaced with $\rmj\ul{E}/\omega$ and $\rmj\ul{H}/\omega$, respectively. For unbounded plane waves, both $\ul{M}_s$ and $\ul{M}_o$ are purely imaginary. 

\section{Average LM and AM of Random Fields}
\subsection{Arbitrary Distance or Frequency}
To calculate (\ref{eq:avgS}) and (\ref{eq:avgM}) explicitly, we employ the angular spectral plane-wave expansion of random fields \cite{dunn1990}, \cite{arnaPREmar2006}, \cite{hill1998}
\bea
\ul{E} (\ul{r}) = \frac{1}{\Omega}\int\int_\Omega \ul{\cal E}(\Omega) \exp (-\rmj \ul{k} \cdot \ul{r}) \rmd\Omega
\label{eq:Ealpha}
\eea
with a similar expansion for $\ul{H}(\ul{r})$,
leading to
\bea
E_\alpha(\ul{r}) H^*_\beta(\ul{r}) &=& \frac{1}{\Omega} \left ( \int\int_\Omega \right )^2 (\ul{\cal E}_1(\Omega_1) \cdot\ul{1}_\alpha) (\ul{\cal H}^*_2(\Omega_2) \cdot\ul{1}_\beta) \nonumber\\
&~& \times \exp [-\rmj (\ul{k}_1 - \ul{k}^{*}_2)\cdot\ul{r}] \, \delta(\Omega_1{\Delta}\Omega_2) \rmd\Omega_1 \rmd\Omega_2
\nonumber
\label{eq:EalphaHbeta}\\
\eea
for $\alpha,\beta \in \{x,y,z\}$. Here, $\Omega = 2\pi$ sr is the solid angle of the half space of incidence above the boundary ($z\geq 0$);  
$\Omega_1 \Delta \Omega_2 \stackrel{\Delta}{=} (\Omega_1\cup\Omega_2)\setminus(\Omega_1\cap\Omega_2)$ for $\Omega_1$, and $\Omega_2$ as point sets; 
$\delta(\cdot)$ is Kronecker's delta ($\delta(\emptyset)=1$, and 0 otherwise);
$\rmd\Omega_i = \sin \theta_i \rmd\theta_i \rmd\phi_i$ 
with elevation angle $\theta_i$ and azimuth angle $\phi_i$ in standard spherical coordinates ($0 < \theta_i \leq \pi/2$, $0 < \phi_i \leq 2\pi$).

For each plane-wave component $\{\ul{\cE}_i,\ul{\cH}_i,\ul{k}_i\}$, a TE/TM decomposition is performed with respect to its plane of incidence $o k_i z$ defining $\phi_i=0$ \cite{dunn1990}, \cite{arnaPREmar2006}. 
The incident plus reflected field is aggregated across the angular spectrum by integration across $\phi_i$, $\theta_i$ and the uniformly distributed polarization angle $-\psi_i$ in the locally transverse plane.
For example, for $\alpha=x$ and $\beta=y$, substituting \cite[eqs. (10)--(15)]{arnaPREmar2006} into (\ref{eq:EalphaHbeta}) yields
\bea
&~& \hspace{-0.8cm} E_x(z) H^*_y(z)
=
\frac{\rmj}{\pi} 
\int^{\pi/2}_0 \cos \theta_1 \sin(2 k_1 z \cos \theta_1) \sin \theta_1 \rmd\theta_1
\nonumber\\
&~& \times \int^{2\pi}_0 \left ( {\cal E}_{1\theta} {\cal H}^*_{2\phi} \cos^2 \phi_1 - {\cal E}_{1\phi} {\cal H}^*_{2\theta} \sin^2 \phi_1 \right ) \rmd\phi_1 
\label{eq:ExHyIntegralExpression}
\eea
with the locally transverse $\ul{\cE}_i$ and $\ul{\cH}_i$ ($i=1,2$) given by
\cite{arnaPREmar2006}
\bea
\cE_{i\phi} = {\cal E}_0 \cos \psi_i,~
\cE_{i\theta} = - {\cal E}_0 \sin \psi_i,~
\cE_{ik} = 0
\label{eq:EthetaEphi}\\
\cH_{i\phi} = \frac{{\cal E}_0}{\eta_0} \sin \psi_i,~
\cH_{i\theta} = \frac{{\cal E}_0}{\eta_0} \cos \psi_i,~
\cH_{ik} = 0
\label{eq:HthetaHphi}
\eea
where ${\cal E}_0 \equiv {\cal E}^\prime_0 - \rmj {\cal E}^{\prime\prime}_0$ 
is the complex amplitude of the circular %\footnote{Here we consider the incident $\cE_0$ to be randomly polarized ($\langle\ul{M}_s\rangle = \ul{0}$) and complex circular. Then, because of additivity in the SAM/OAM decomposition (\ref{eq:Ms_plus_Mo}), $\langle\ul{M}\rangle = \langle\ul{M}_o\rangle$.} 
electric field of each plane-wave component, with $\langle |\cE_0|^2 \rangle = 2 \langle \cE^{\prime(\prime)^2}_0 \rangle = \langle |\ul{E}|^2\rangle/4$ and $\eta_0 = \sqrt{\mu_0/\eps_0}$.
Integration of (\ref{eq:ExHyIntegralExpression}) followed by ensemble averaging (denoted as $\langle \cdot \rangle$) over ${\cal E}_0$ and $\psi$ 
yields
\bea
\langle E_x H^*_y \rangle = - \langle E_y H^*_x \rangle = 
- \rmj \frac{%2\pi 
\langle |{\cal E}_0|^2 \rangle}{\eta_0} j_1 (2 k z)
\label{eq:avgExHy}
\eea
where here and for later use 
\bea
j_0 (2kz) &\stackrel{\Delta}{=}& {\sinc(2kz)}
\label{eq:def_j0}\\
j_1 (2 k z) &\stackrel{\Delta}{=}& 
\frac{\sin ( 2 k z )}{( 2 k z )^2} 
- \frac{\cos ( 2 k z )}{2 k z} = -j^\prime_0(2kz)
\label{eq:def_j1}\\
j_2(2kz) &\stackrel{\Delta}{=}& \left ( \frac{3}{(2kz)^3} - \frac{1}{2kz}\right ) \sin(2kz) - \frac{3\cos(2kz)}{(2kz)^2}~~~~
\label{eq:def_j2} \\
%\nonumber\\
j_3(2kz) &\stackrel{\Delta}{=} & \left ( \frac{15}{(2kz)^4} - \frac{6}{(2kz)^2}\right ) \sin(2kz) \nonumber\\
&~& - \left ( \frac{15}{(2kz)^3} - \frac{1}{2kz} \right ) \cos(2kz)
\label{eq:def_j3_bis}
\eea
are spherical Bessel functions of the first kind and order zero to three, respectively.
With an analogous calculation, 
\bea
\langle E_x H^*_z \rangle = \langle E_z H^*_x \rangle =  
%\nonumber\\&~& 
\langle E_y H^*_z \rangle = \langle E_z H^*_y \rangle = 0
\label{eq:avgother}
\eea
because their kernel's azimuthal dependence is of the form $\sin\phi$ or $\cos\phi$, as opposed to their square in (\ref{eq:ExHyIntegralExpression}).
Combining (\ref{eq:avgM}), (\ref{eq:avgExHy}) and (\ref{eq:avgother}) yields the ensemble averaged LM and AM as
\bea
\langle \ul{P} (\ul{r},\omega) \rangle \equiv \frac{\langle \ul{S} \rangle}{\rmc^2 }
&=& - \rmj \frac{ 
\langle |{\cal E}_0|^2 \rangle}{\rmc^2 \eta_0} j_1 (2 k z) ~ \ul{1}_z 
\label{eq:finalavgS}\\
\langle \ul{M} (\ul{r},\ul{0},\omega) \rangle &=& - \rmj \frac{ 
\langle |{\cal E}_0|^2 \rangle}{\rmc^2 \eta_0} j_1 (2 k z) 
\, \left ( y \ul{1}_x  - x \ul{1}_y \right )~~~~~
\label{eq:finalavgM}
\eea
i.e., $\langle \ul{M} \rangle = \ul{r} \times \langle \ul{P} \rangle = - r \langle P \rangle \ul{1}_\phi$.
The result for $\langle \ul{M} (\ul{r},\ul{0},\omega) \rangle$ in (\ref{eq:finalavgM}) extends to general $\langle\ul{M}(\ul{r}, \ul{r}_0, \omega ) \rangle$ by subtracting $\ul{r}_0 \times \langle \ul{P}(\ul{r},\omega) \rangle $ from $\langle \ul{M} (\ul{r},\ul{0},\omega) \rangle$.

The zero real parts of the power flux densities $\langle {S} \rangle = \rmc^2\langle {P} \rangle$ and $\rmc^2\langle {M} \rangle$  indicate zero time-averaged energy flow in normal (i.e., $\ul{1}_z$-directed, longitudinal) and transverse azimuthal directions, respectively. Thus, the ensemble averaged incident power and the propagating power after reflection off a PEC boundary cancel.
The combined LM and AM flux densities depend on all three spatial coordinates via $j_1(2kz)$ and the radial transverse distance $||\ul{r} \times \ul{1}_z|| = \sqrt{x^2+y^2}$. At any height $kz$ above the boundary, $\langle \ul{M} \rangle$ is tangential ($\langle M_z \rangle = 0$) and purely solenoidal ($\ul{\nabla} \times \langle\ul{M} \rangle \not = \ul{0}$, $\ul{\nabla} \cdot \langle \ul{M} \rangle = 0$).
Similar to deterministic fields in free space, $\langle\ul{M}\rangle$ is complementary to the irrotational and normal $\langle \ul{P} \rangle$, i.e., $\ul{\nabla} \times \langle\ul{P} \rangle = \ul{0}$, $\ul{\nabla} \cdot \langle \ul{P} \rangle \not = 0$.
In summary, for random fields near a planar PEC boundary, the combined linear (longitudinal) and angular (azimuthal) average power fluxes are reactive and characterized by a 3-D hybrid vector
\bea
\langle \ul{{\mit\Pi}}(\ul{r},\ul{k}) \rangle \stackrel{\Delta}{=} 
 - \rmj \frac{ 
\langle |{\cal E}_0|^2 \rangle}{\eta_0} j_1 (2 k z) \left ( y \ul{1}_x  - x \ul{1}_y
 + \ul{1}_z \right ).
\label{eq:finalavgSandM}
\eea
Its LM and AM components can be measured, e.g., using a modified magic T or a turnstile junction with L-shaped in-plane sections that contain inductive (diaphragm) shunt loads.
Alternatively, helical and planar chiral power sensors with reactive loading may be used to measure the longitudinal and transverse progression of the  phase for SAM and OAM, respectively.

\begin{figure}[htb] \begin{center} \begin{tabular}{c}
\vspace{-0.5cm}\\
\hspace{-0.7cm}
\includegraphics[scale=0.50]{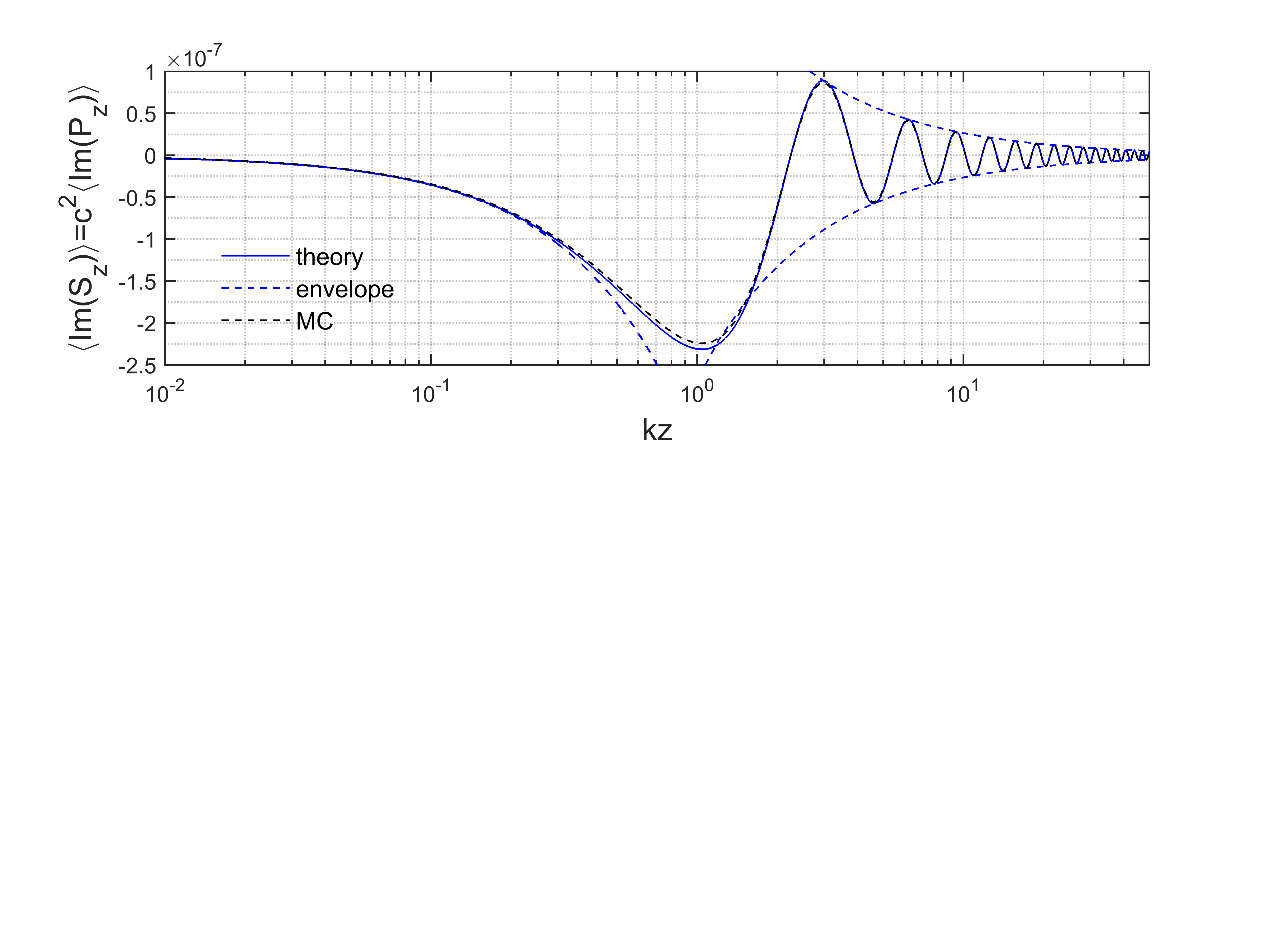}\\
\vspace{-4.5cm}\\
\end{tabular}
\end{center}
{
\caption{\label{fig:avgSz} \small
Amplitude of mean linear or angular power flux density components $\langle S_z \rangle = {\rmc}^2 \langle P_z \rangle = {\rmc}^2 \langle M_x \rangle / y =  - {\rmc}^2 \langle M_y \rangle / x$ [in units W/m$^2$] as a function of height $kz$ above a PEC boundary for $\langle{{\cal E}^\prime_0}^2 \rangle = \langle{{\cal E}^{\prime\prime}_0}^2 \rangle = 10^{-4}$ (V/m)$^2$. 
Blue solid: theory [eq. (\ref{eq:finalavgS})];  blue dashed: envelopes ${\mit\Xi}^\pm_s$ for $kz\gg 1/2$ [eq. (\ref{eq:asympenvelopeS_HF})] and for $kz\ll \sqrt{10}$ [eq. (\ref{eq:asympenvelopeS_LF})]; 
black dashed: MC simulation.
}
}
\end{figure}

The results (\ref{eq:finalavgS})--(\ref{eq:finalavgM}) are verified using a Monte Carlo (MC) simulation of the random plane-wave spectrum (\ref{eq:Ealpha}).
We used 370 values of $kz$ ranging from 0.01 to 50 in logarithmic steps of 0.01 and define $x=y=1$ m. For each $kz$, a set of $n_\theta \times n_\phi \times n_\psi = 32 \times 16\times 16$ uniformly spaced angles of incidence and polarization were generated across $(0,\pi/2] \times (0,2\pi]\times (0,2\pi]$ with $n=30$ complex random fields $\cE_0$ for each $\ul{k}$. This yields $245\/760$ plane waves per $kz$.
Fig. \ref{fig:avgSz} shows the resulting average LM power density $\langle S(kz) \rangle$ and, by extension, the transverse components of the AM power density $\rmc^2\langle M(kz)\rangle$ normalized by the values of $x$ and $y$. 

%**************FIGURE 600****************
%**************FIGURE ****************

\subsection{Optimum Locations of Field vs. Power Sensors\label{sec:optimum}}
On and at asymptotically large distances from the boundary $(kz \rightarrow +\infty)$, $\langle \ul{P} \rangle$ and $\langle \ul{M} \rangle$ vanish, which confirms the results for unbounded random fields \cite[eqs. (19)-(20)]{paololegacy}, \cite[eq. (51)]{hill1998}.
They also vanish for $j_1(2kz)=0$, i.e., at frequencies and distances that are related as
\bea 
\tan(2kz_s) = 2kz_s
\label{eq:cond_S}
\eea 
viz., at $z_s=0+$ or $z_s \simeq (2m+3)\pi/(4k)$ when $kz_s \gg 1$ ($m=0,1,2,\ldots$).
Such measurement locations are preferential when aiming to avoid the influence of a PEC boundary on the average LM and AM using a power sensor.
Note that these locations $z_s$ apply strictly to dot sensors and CW excitation, and vice versa. Inevitably, the optima become blurred for a sensor of finite size or for nonzero bandwidths owing to local averaging.

These findings for $\langle {P} \rangle$ and $\langle {M} \rangle$ complement those for the average energy densities $\langle U_e \rangle$ and $\langle U_m \rangle$ of the 3-D total electric or magnetic vector field (no subscript), the 2-D tangential (subscript $t$), and the 1-D Cartesian components ($x,y,z$) \cite{dunn1990}, \cite{arnaTEMCmay2006}, written here in an alternative but equivalent form as
\bea
\langle U (kz) \rangle 
&=& \eps_0 \langle{|\cal E}_0|^2 \rangle \left [1 \mp \frac{j_0(2kz) }{3} \pm \frac{2 j_2(2kz)}{3} \right ]
~~~~\label{eq:avgUeUm}\\
\langle U_{t} (kz) \rangle 
&=& \frac{2\eps_0 \langle{|\cal E}_0|^2 \rangle}{3} \left [ 1 \mp j_0(2kz) \pm \frac{j_2(2kz)}{2} \right ]
~~~\label{eq:avgUetUmt}\\
\langle U_{(x)(y)} (kz) \rangle 
&=& \frac{\eps_0 \langle{|\cal E}_0|^2 \rangle}{3} \left [ 1 \mp j_0(2kz) \pm \frac{j_2(2kz)}{2} \right ]
\label{eq:avgUexyUmxy}\\
\langle U_{z} (kz) \rangle 
&=& \frac{\eps_0 \langle{|\cal E}_0|^2 \rangle}{3} \left [ 1 \pm j_0(2kz) \pm j_2(2kz) \right ]
\label{eq:avgUezUmz}
\eea 
where upper and lower signs apply to electric $(U=U_{e})$ and magnetic $(U=U_{m})$ densities, respectively. 
Unlike (\ref{eq:cond_S}) for $\langle {S} \rangle$, the asymptotic values of $\langle U \rangle$ for $kz \rightarrow \infty$ are now reached for $j_0(2kz) = 2 j_2(2kz)$, i.e., when
\bea
\tan(2kz_u) = \frac{2kz_u}{1-2(kz_u)^2},~~~(kz_u \not = 0) \label{eq:cond_U_e}
\eea
viz., at $z_u \simeq (m+1)\pi/(2k)$ for $kz_u \gg 1$.
The solutions of (\ref{eq:cond_S}) and (\ref{eq:cond_U_e}) are separated by $kz_u-kz_s\simeq\pi/4$ for $m\gg 1$. 

Similarly, the frequencies and locations for reaching the asymptotic 2-D tangential energy $\langle U_{t} (kz \rightarrow \infty) \rangle$ follow from (\ref{eq:avgUetUmt}) as solutions of $j_0(2kz) = j_2(2kz)/2$, i.e., when
\bea
\tan(2kz_{u,t}) = \frac{2kz_{u,t}}{1-(2kz_{u,t})^2},~~~(kz_{u,t} \not = 0) \label{eq:cond_U_et}
\eea
viz., at $z_{u,t} \simeq (m+1)\pi/(2k)$ when $kz_{u,t} \gg 1$.
For the 1-D tangential components $\langle U_{x,y} (kz \rightarrow \infty) \rangle$, the solutions follow from (\ref{eq:avgUexyUmxy}) as being identical to those for $\langle U_{t}\rangle$.
Finally, for the 1-D normal component $\langle U_{z} (kz \rightarrow \infty) \rangle$, the locations are those for $\langle S \rangle$ but exclude the boundary plane:
from (\ref{eq:avgUezUmz}), $j_0(2kz)=-j_2(2kz)$ is satisfied when
\bea 
\tan(2kz_{u,z}) = 2kz_{u,z},~~~(kz_{u,z} \not = 0)\label{eq:cond_UenEmn}
\eea 
viz., at $z_{u,z} \simeq (2m+3)\pi/(4k)$ when  $kz_{u,z} \gg 1$.

The practical significance of these different optimum values is that, depending on whether one measures either the Cartesian or vector field energy density (or intensity) using a field sensor (wire or loop probe) or the reactive LM or AM power flux using a power sensor (aperture antenna), these devices should be placed at different heights above a PEC boundary in order to eliminate the effect of the boundary on the measurement.

\begin{figure}[htb] \begin{center} \begin{tabular}{c}
\vspace{-1cm}\\
\hspace{-0.75cm}
\includegraphics[scale=0.50]{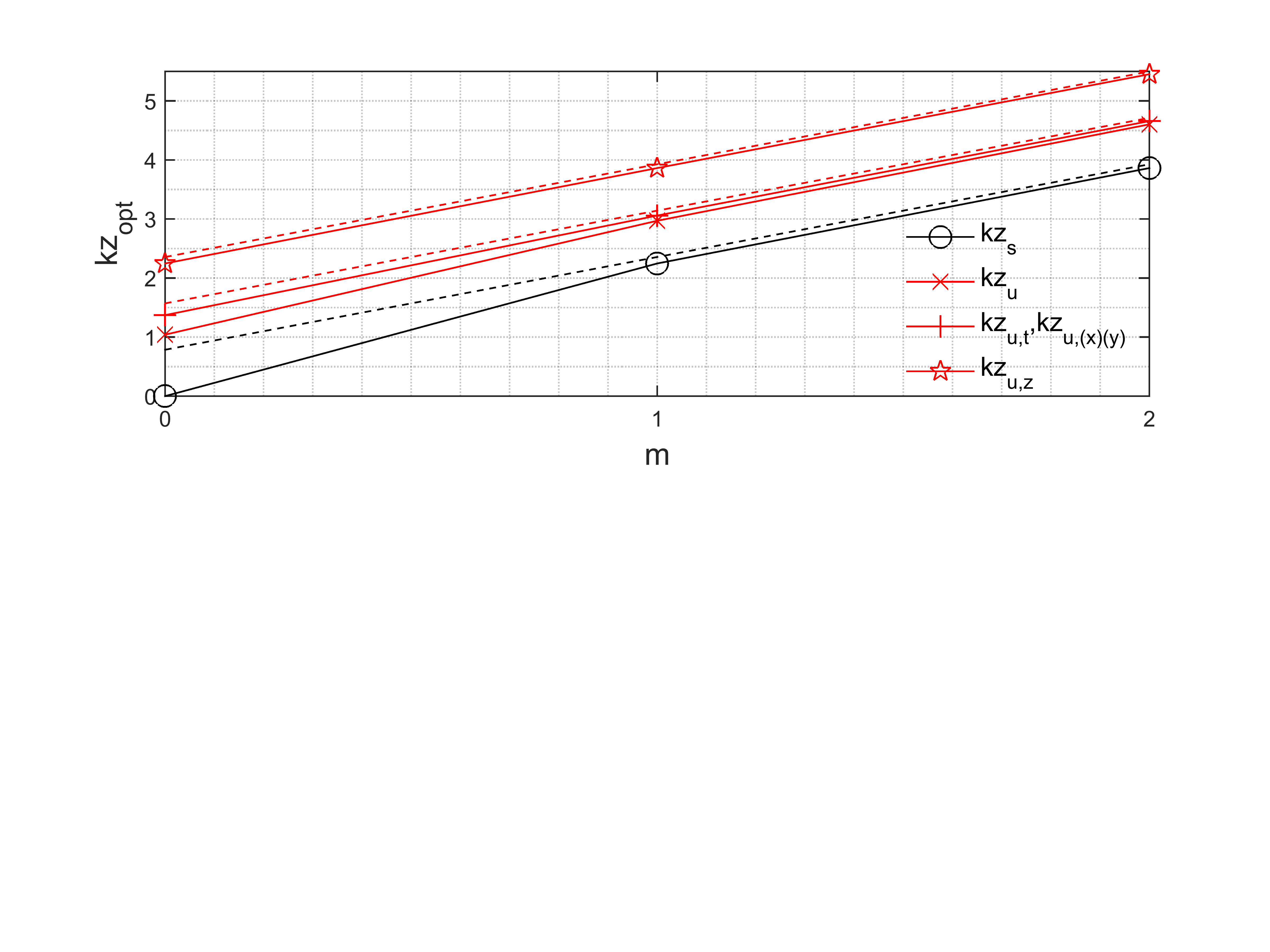}\\
\vspace{-4.5cm}\\
\end{tabular}
\end{center}
{
\caption{\label{fig:optimumloc} \small
First three optimum locations $kz_{\rmopt}(m=0,1,2)$ for a reactive power sensor (black) vs. various field sensors (red) near a PEC boundary. Symbols: actual optima (cf. Tbl. \ref{tbl:solzeroes}), dashed lines: asymptotic laws.
}
}
\end{figure}

The first few optimum locations for $kz_s$, $kz_u$, $kz_{u,t}\equiv kz_{u,(x)(y)}$ and $kz_{u,z}$ are listed in Tbl. \ref{tbl:solzeroes} and shown in Fig. \ref{fig:optimumloc}. 
In each case, the optimal distances are spaced by asymptotically $\pi/2$, as follows from the asymptotic approximation \cite{arf}
\bea
j_\ell(2kz) \simeq \frac{\sin(2kz-\ell\pi/2)}{2kz}~~~{\rm for}~kz\gg \frac{\ell(\ell+1)}{4}.
\label{eq:asympBessel_HF}
\eea 
Excluding the boundary and DC regime ($kz\not =0$), the shortest optimum distance ($m=0$) is attained for a 3-D isotropic field sensor ($u$), followed by a 1-D  ($u_x$ or $u_y$) or 2-D ($u_t$) tangential field probe. 
A normal field sensor ($u_z$) and power sensor ($s$) must be placed farthest, at more than twice the distance for the isotropic field probe.
If the boundary plane is included, however, then power sensors exhibit the shortest (viz., zero) optimum distance.
\begin{table}[!htb]
\begin{center}
\begin{tabular}{||l||l|l|l|l|l|l||}
\hline
\hline
$m$ & 0 & 1 & 2 & 3 & 4 & 5\\
\hline
\hline
$kz_s$ & 0 & 2.247 & 3.863 & 5.452 & 7.033 & 8.610\\
\hline
$kz_u$ & 1.041 & 2.970 & 4.603 & 6.202 & 7.790 & 9.371\\
\hline
$kz_{u,(t)(x)(y)}$ & 1.372 & 3.058 & 4.658 & 6.242 & 7.821 & 9.398\\
\hline
$kz_{u,z}$ &  2.247 & 3.863 & 5.452 & 7.033 & 8.610 & 10.19\\
\hline
\hline
\end{tabular}
\end{center}
\caption{\label{tbl:solzeroes} \small First six locations $kz_s$ and $kz_u$ for optimal placement of power or field sensors with various orientations.}
\end{table} 

%**************FIGURE 61****************
%**************FIGURE ****************

Note that for field intensity (energy) sensors, the optimal $kz$ values are irrespective of the electric or magnetic type of the sensor.
On the other hand, for a combined electric-magnetic sensor or an (un)intentional receptor that is sensitive to different 
%electric and magnetic field 
components in different measures and orientations, its optimal distance or frequency can be estimated on a case by case basis by decomposing its orientation into normal and tangential components, which provide the weights for an appropriate superposition of the components of its average response based on (\ref{eq:finalavgS}) and (\ref{eq:avgUeUm})--(\ref{eq:avgUexyUmxy}).
 
\subsection{Maximum Deviations of Mean Power and Energy} 
\subsubsection{Envelopes\label{sec:env}}
The results in Sec. \ref{sec:optimum} listed optimum locations $z_{\rmopt}$ for a specified frequency $k$, or vice versa.
In EMC practice, signals are often wideband or being swept across frequency.
In addition, surface imperfections, proximity of scatterers or other surfaces, etc., may perturb the field and affect the optimality of these locations. 
Given these spatio-spectral uncertainty or aggregation effects, a more distant optimum ($kz_{\rmopt} \gg 1$) where the fluctuations of $\langle S(kz)\rangle$ are weaker may actually be preferential, in order to reduce the sensitivity on the optimum placement.
The envelopes of $\langle S(kz)\rangle$ then represent the maximum deviation from the asymptotic value $\langle S(kz \rightarrow +\infty)\rangle$ that may be expected as a suboptimal alternative.
 
\begin{figure}[htb] \begin{center}
\begin{tabular}{c}
\vspace{-0.8cm}\\
\hspace{-0.75cm}
\includegraphics[scale=0.50]{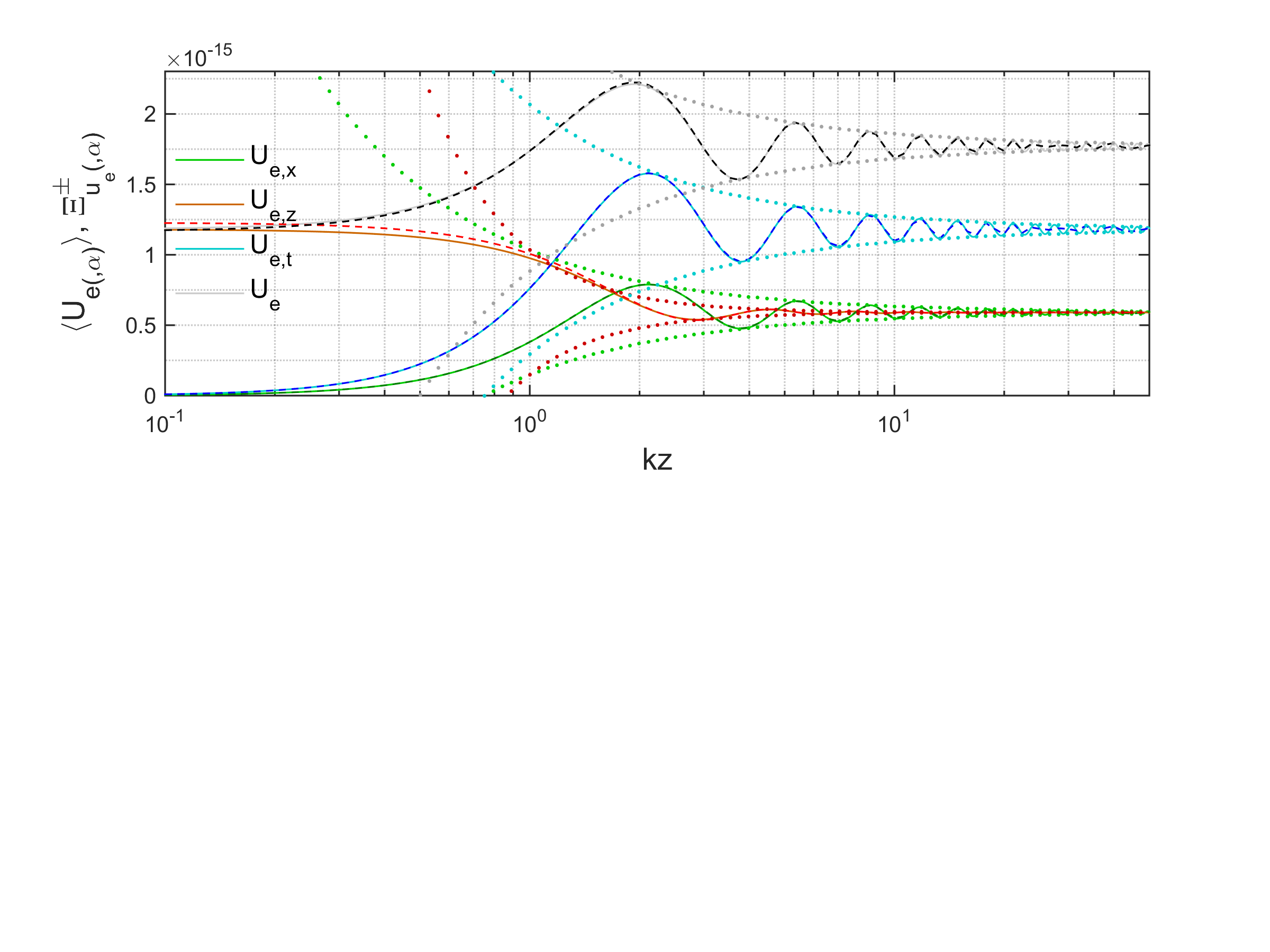}\\
\vspace{-4.2cm}\\
(a)\\
\hspace{-0.8cm}
\includegraphics[scale=0.50]{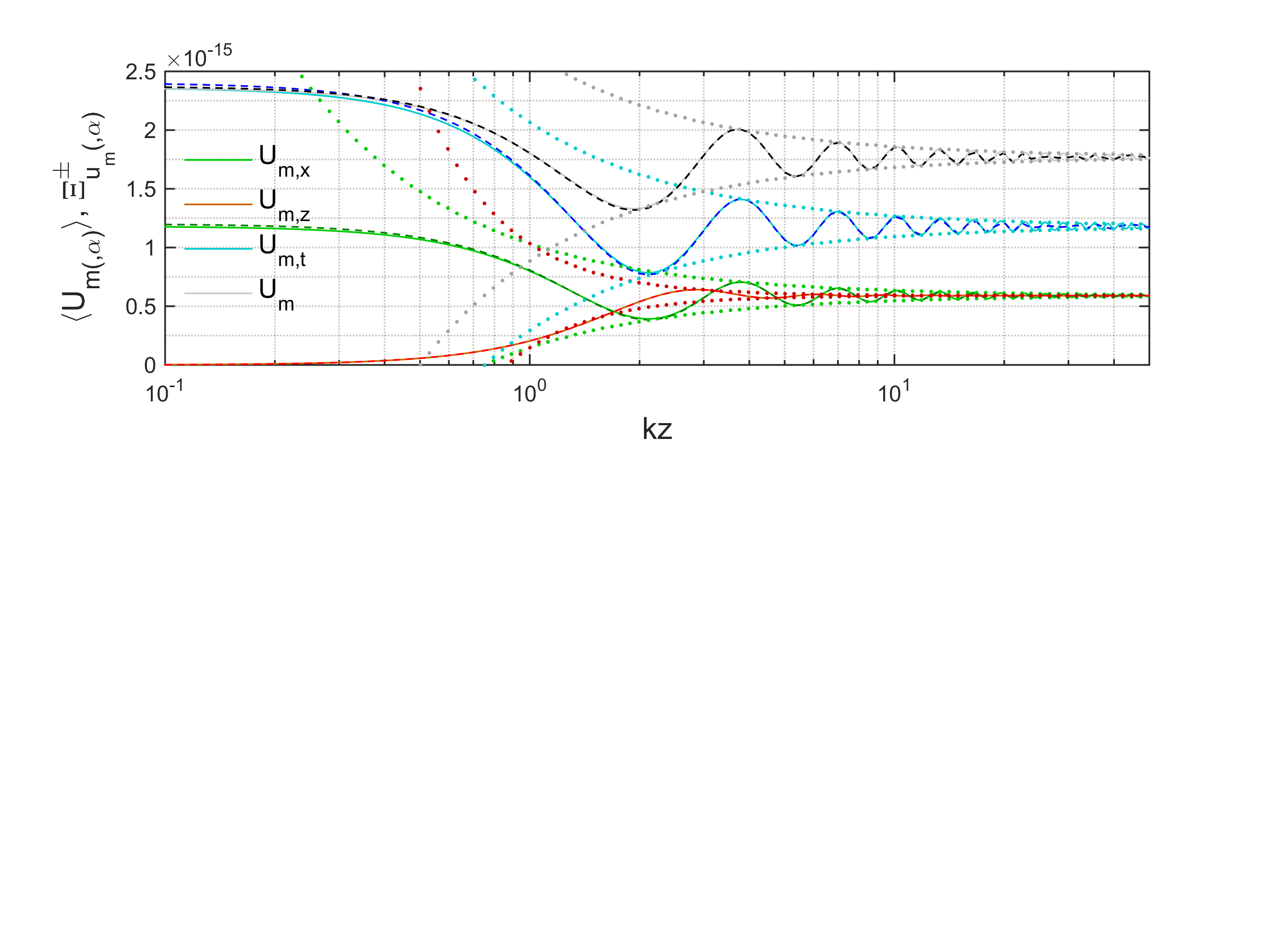}\\
\vspace{-4.2cm}\\
(b)
\\
\end{tabular}
\end{center}
{
\caption{\label{fig:Uenv_all} \small
Theoretical averages [eqs. (\ref{eq:avgUeUm})--(\ref{eq:avgUezUmz})] (light solid), their MC simulations (dark dashed), and asymptotic envelopes ${\mit\Xi}^\pm$ [eqs. (\ref{eq:envU})--(\ref{eq:envUz})] (light dotted), for (a) electric and (b) magnetic energy densities 
[in units J/m$^3$] as a function of height $kz$ above a PEC boundary for $\langle{{\cal E}^\prime_0}^2 \rangle = \langle{{\cal E}^{\prime\prime}_0}^2 \rangle = 10^{-4}$ (V/m)$^2$. Black/grey: $\langle U \rangle$; 
blue/cyan: $\langle U_t \rangle$;
maroon/red: $\langle U_z \rangle$;
olive/green: $\langle U_x \rangle$. 
}
}
\end{figure}

The asymptotic analytic representation  
$\tilde{\langle S\rangle} \stackrel{\Delta}{=} \langle S\rangle +\rmj {\cal H}[\langle S \rangle]$ follows with the Hilbert transformation ${\cal H}[j_1(2kz)] \simeq \cos(2kz-\pi/2)/(2kz)$ for (\ref{eq:asympBessel_HF}) and $-\rmj \equiv \exp(-\rmj \pi/2)$ as
\bea
\langle \widetilde{S}(kz\gg 1/2) \rangle 
\simeq \frac{\langle |\cE_0|^2\rangle}{2\eta_0 kz} \exp \left [ \rmj ( 2kz - \pi ) \right ] 
\eea 
whose signed magnitude defines the upper and lower envelopes
\bea
{\mit\Xi}^\pm_s(kz \gg 1/2) \simeq \pm \frac{ 
\langle |{\cal E}_0|^2 \rangle}{2 \eta_0 kz}.
\label{eq:asympenvelopeS_HF}
\eea
These are indicated in Fig. \ref{fig:avgSz}.
Their half separation $\Delta {\mit\Xi}_s(kz) \stackrel{\Delta}{=} [{\mit\Xi}^+_s(kz)-{\mit\Xi}^-_s(kz)]/2$ measures the maximum absolute deviation from the asymptotic mean power $\langle S(kz\rightarrow +\infty)\rangle = [{\mit\Xi}^+_s(kz)+{\mit\Xi}^-_s(kz)]/2 = 0$.
For {\em both\/} electric and magnetic energy densities, similar definitions and calculations\footnote{In (\ref{eq:envUz}), the envelopes of $\langle U_z \rangle$ are of leading second order $(kz)^{-2}$, because of cancelling first-order terms. For the other energies, additional second- and higher-order terms in $(kz)^{-1}$ provide corrections when $kz \not \gg 3/2$.} yield
\bea
{\mit\Xi}^\pm_{u}(kz\gg 3/2) 
&=& 
\eps_0 \langle{|\cal E}_0|^2 \rangle \left (
1 \pm \frac{1}{2kz} \right ) %,~~~kz \gg 3/2
\label{eq:envU}\\
{\mit\Xi}^\pm_{u,(x)(y)}(kz \gg 3/2) 
&=& 
\frac{\eps_0 \langle{|\cal E}_0|^2 \rangle}{3} \left ( 1 \pm \frac{3}{4kz} \right ) \nonumber \\ %,~~~kz \gg 3/2
&=&
\frac{1}{2} {\mit\Xi}^\pm_{u,t}(kz\gg 3/2)
\label{eq:envUt}
\label{eq:envUx}\\
{\mit\Xi}^\pm_{u,z}(kz\gg 3/2) 
&=& 
\frac{\eps_0 \langle{|\cal E}_0|^2 \rangle}{3} 
\left ( 1 \pm \frac{3}{4(kz)^2} \right ) %,~~~kz \gg 3/2~~~
\label{eq:envUz}
\eea
where upper and lower signs now correspond to upper and lower envelopes.
These are shown in Fig. \ref{fig:Uenv_all} together with (\ref{eq:avgUeUm})--(\ref{eq:avgUezUmz}).

%**************FIGURE 1 (\matlab6p1\work\OAM_SD\plot_OAM_envelopes_v2.m)****************
%**************FIGURE 1****************

The associated maximum relative deviations are
\bea
\Delta {\xi}_u(kz \gg 3/2) &\stackrel{\Delta}{=}& 
\frac{\Delta {\mit\Xi}_u(kz \gg 3/2)}{{\mit\Xi}_u(kz\rightarrow +\infty)} 
= \frac{1}{2kz}\label{eq:deltav_u}\\
\Delta {\xi}_{u,(x)(y)}(kz \gg 3/2) &=& \Delta {\xi}_{u,t}(kz \gg 3/2) 
= \frac{3}{4kz}~~~\label{eq:deltav_u_t}
\label{eq:deltav_u_xy}\\
\Delta {\xi}_{u,z}(kz \gg 3/2) 
&=& \frac{3}{4(kz)^2}
\label{eq:deltav_u_z}
\eea 
For example, at $kz=10$ we find
$\langle U_{x}(10)\rangle =0.3104\,\eps_0 \langle |{\cal E}_0|^2 \rangle$, whereas
${\mit\Xi}^\pm_{u,x}(10)=[(1\pm 0.075)/3]\,\eps_0 \langle{|\cal E}_0|^2\rangle$ results in $\Delta {\mit\Xi}_{u,x}(10) = 0.025\,\eps_0 \langle{|\cal E}_0|^2 \rangle$ and  $\Delta {\xi}_{u,x}(10) = 0.075$.
For $\langle S \rangle$, a different normalization is adopted than for $\langle U\rangle$, viz., 
\bea
\Delta {\xi}_{s}(kz \gg 1/2) 
\stackrel{\Delta}{=} 
\frac{\Delta {\mit\Xi}_s(kz \gg 1/2)}{\langle |\cE_0|^2\rangle/\eta_0}
=
\frac{1}{2kz} 
\label{eq:deltav_s}
\eea 
in view of ${\mit\Xi}_s(kz\rightarrow+\infty)=0$.
The motivation for this particular choice 
will become clear in Sec. \ref{sec:WV} and corresponds to a normalization of $\Delta {\mit\Xi}_{u}$ and $\langle U \rangle$ by $\eps_0 \langle |\cE_0|^2\rangle$.

\subsubsection{Maximum Deviation\label{sec:maxdev}}
The above relative deviations can be used with a specified minimum distance of the sensor to the PEC boundary to estimate the maximum local deviation from the asymptotic mean value, or vice versa.    
Tbl. \ref{tbl:swing} shows these deviations for $kz=\pi/2$, $\pi$ and $2\pi$.
The entries indicate that a normally directed field sensor gives a smaller maximum deviation than a power sensor, at sufficiently large distances. For example, $\Delta {\xi}_{u,z}=7.60\%$ compared to $\Delta {\xi}_s=15.9\%$ at $z=\lambda/2$ with the chosen normalization, whereas at $z=\lambda/4$ the maximum deviations of the normalized $\langle U_z \rangle$ and $\langle S_z \rangle$ are more similar ($30.4\%$ vs. $31.8\%$, respectively).
\begin{table}[!htb]
\begin{center}
\begin{tabular}{||l||l|l|l|l||}
\hline
\hline
$kz$    & $\Delta {\xi}_s$ & $\Delta {\xi}_u$ & $\Delta {\xi}_{u,(t)(x)(y)}$ & $\Delta {\xi}_{u,z}$\\
\hline
\hline
$\pi/2$ & 0.318 & 0.318 & 0.477 & 0.304\\
\hline
$\pi$ & 0.159 & 0.159 & 0.239 & 0.076\\
\hline
$2\pi$ & 0.080 & 0.080 & 0.119 & 0.019\\
\hline
\hline
\end{tabular}
\end{center}
\caption{\label{tbl:swing} \small Maximum relative deviation for envelopes of $\langle S_z \rangle / (\langle |\cE_0|^2\rangle/\eta_0)$ and $\langle U_\alpha \rangle / (\eps_0\langle |\cE_0|^2\rangle)$ at quarter-, half-, and full-wavelength distances from a PEC boundary.}
\end{table} 

At the other extreme of very low frequencies and/or small distances from the boundary, using the approximation \cite{arf}
\bea
j_\ell(2kz) \simeq \frac{(2kz)^\ell}{(2\ell+1)!!}~~~{\rm for}~ kz \ll \sqrt{\frac{(2\ell+2)(2\ell+3)}{\ell+1}}
\eea 
the asymptote for quasi-static deviations from $\langle S(kz)\rangle$ is
\bea
{\mit\Xi}^-_s(kz \ll \sqrt{10}) \simeq - \frac{2 
\langle |{\cal E}_0|^2 \rangle kz}{3 \eta_0}
\label{eq:asympenvelopeS_LF}
.
\eea

\subsubsection{Working Volume of a Reverberation Chamber\label{sec:WV}}
\begin{figure}[htb] \begin{center}
\begin{tabular}{c}
\vspace{-0.8cm}\\
\hspace{-0.75cm}
\includegraphics[scale=0.50]{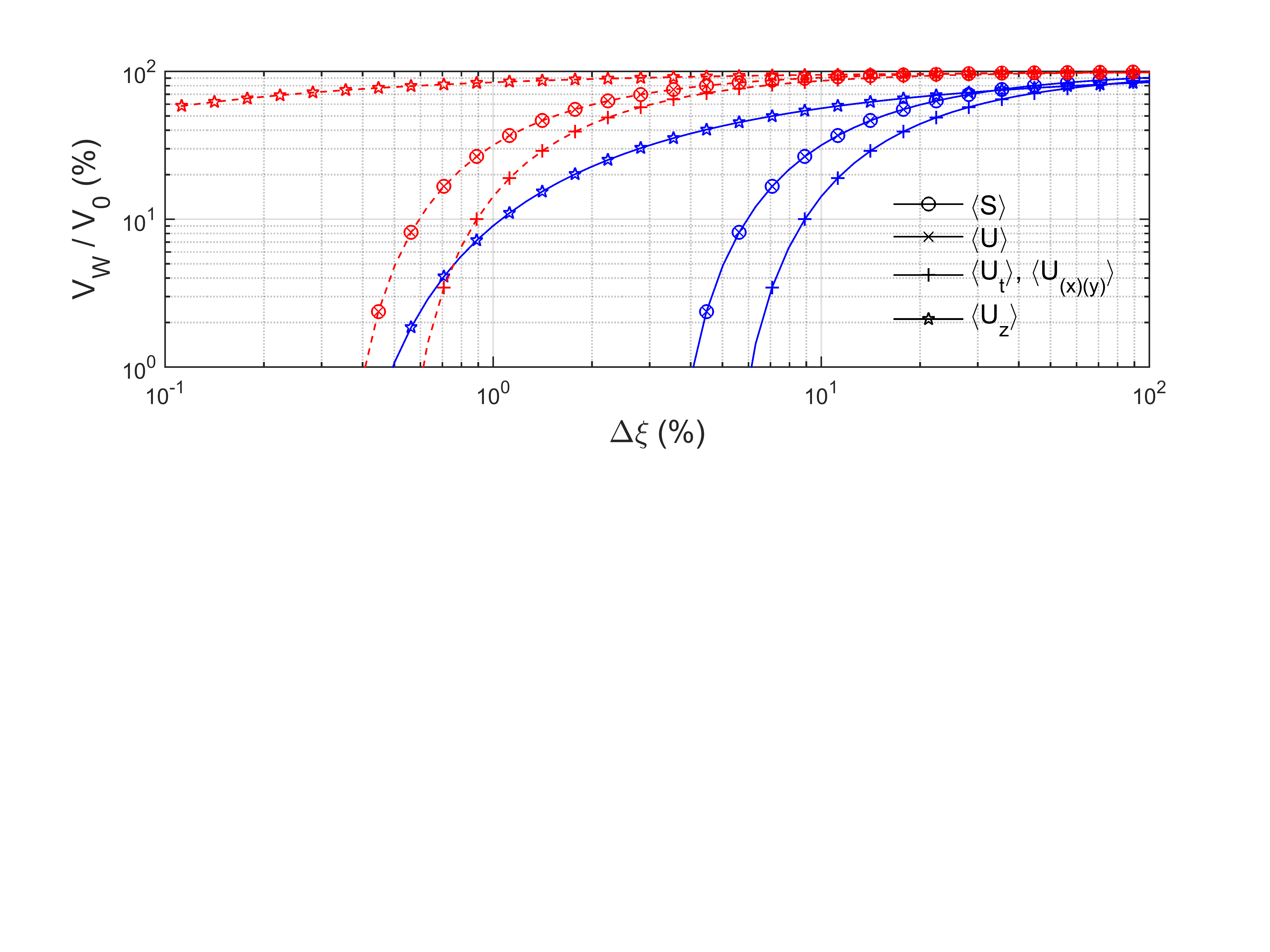}\\
\vspace{-4.5cm}\\
\end{tabular}
\end{center}
{
\caption{\label{fig:WVoverV0ifvXi} \small
Relative working volume $V_W$ as a percentage of cavity volume $V_0$ for specified relative deviation $\Delta{\xi}$ of average power ($S$) or average electric or magnetic energy ($U$).
Blue solid: $V_0=(5\lambda)^3$; red dashed: $V_0=(50\lambda)^3$.
}
}
\end{figure}

The local deviations from the asymptotic $\langle S\rangle$ or $\langle U\rangle$ may be used to estimate the working volume (WV) of a reverberation chamber before measurement.
Consider a cubic cavity of side length $L_0$.
Its WV is a symmetrically located interior cube of sides $L_0 - 2 z_W$ for some $z_W$. On the boundary of this WV, a maximum tolerable relative deviation $\Delta {\xi}$ of the mean energy or power is specified with reference to its asymptotic (ideal) value when $L_0 \rightarrow +\infty$, which yields $kz_W$ from  (\ref{eq:deltav_u})--(\ref{eq:deltav_s}). The relative WV thus follows as 
\bea
\frac{V_W(\Delta {\xi})}{V_0} = 
\left ( \frac{kL_0-2kz_W(\Delta {\xi})}{kL_0} \right )^3.
\label{eq:WV_final}
\eea
This dependence of ${V_W}/{V_0}$ on $\Delta {\xi}$ is shown in Fig. \ref{fig:WVoverV0ifvXi} for average power and energy components, for two cavity size values of $V_0/\lambda^3$.
It is seen that the total (vector) $\langle U_e\rangle$ or $\langle U_m\rangle$ and $\langle S \rangle \equiv \langle S_z \rangle$ yield the same WV.
By contrast, $\langle U_z \rangle$ gives a larger WV, whereas $\langle U_t \rangle$ and $\langle U_{(x)(y)} \rangle$ give a smaller WV. E.g., for $\Delta {\xi} = 0.1$, the value of $V_W/V_0$ based on $\langle U_e \rangle$, $\langle U_m \rangle$ or $\langle S\rangle$ is $31.7\%$, whereas for $\langle U_{(e)(m),z} \rangle$ it is $56.3\%$, whilst for $\langle U_{(e)(m),x} \rangle$ it is merely $14.3\%$.
The larger value of WV for $\langle U_z\rangle$ can be traced to the more rapid decay of its envelope, viz., according to $(kz)^{-2}$, causing the specified uniformity to be reached closer to the boundary.

In practice, adjacent and opposite
walls create additional standing waves, whence
(\ref{eq:WV_final}) serves merely as a first estimate, whose accuracy nevertheless increases with chamber volume $V_0$.
The estimates and results serve as a guide to a more precise evaluation to account for the effects of all boundaries, e.g., based on a full-wave simulation or measurement.

Note that, for a combined electric-magnetic field sensor, $V_W/V_0=1$ for any specified $\Delta {\mit\Xi}$ because the dependence of $\langle U_{e,\alpha}\rangle + \langle U_{m,\alpha}\rangle$ on $kz$ cancels for any component $\alpha$.

%**************FIGURE 1 (\matlab6p1\work\OAM_SD\plot_OAM_envelopes_v2.m)****************
%**************FIGURE 1****************

\section{Conservation of Energy and Momentum}
\subsection{Poynting Theorem for Average Power Flux and Energy\label{sec:Poynting}}
\begin{figure}[htb] \begin{center}
%\vspace{-0.5cm}\\
\hspace{-1.1cm}\\
\includegraphics[scale=0.50]{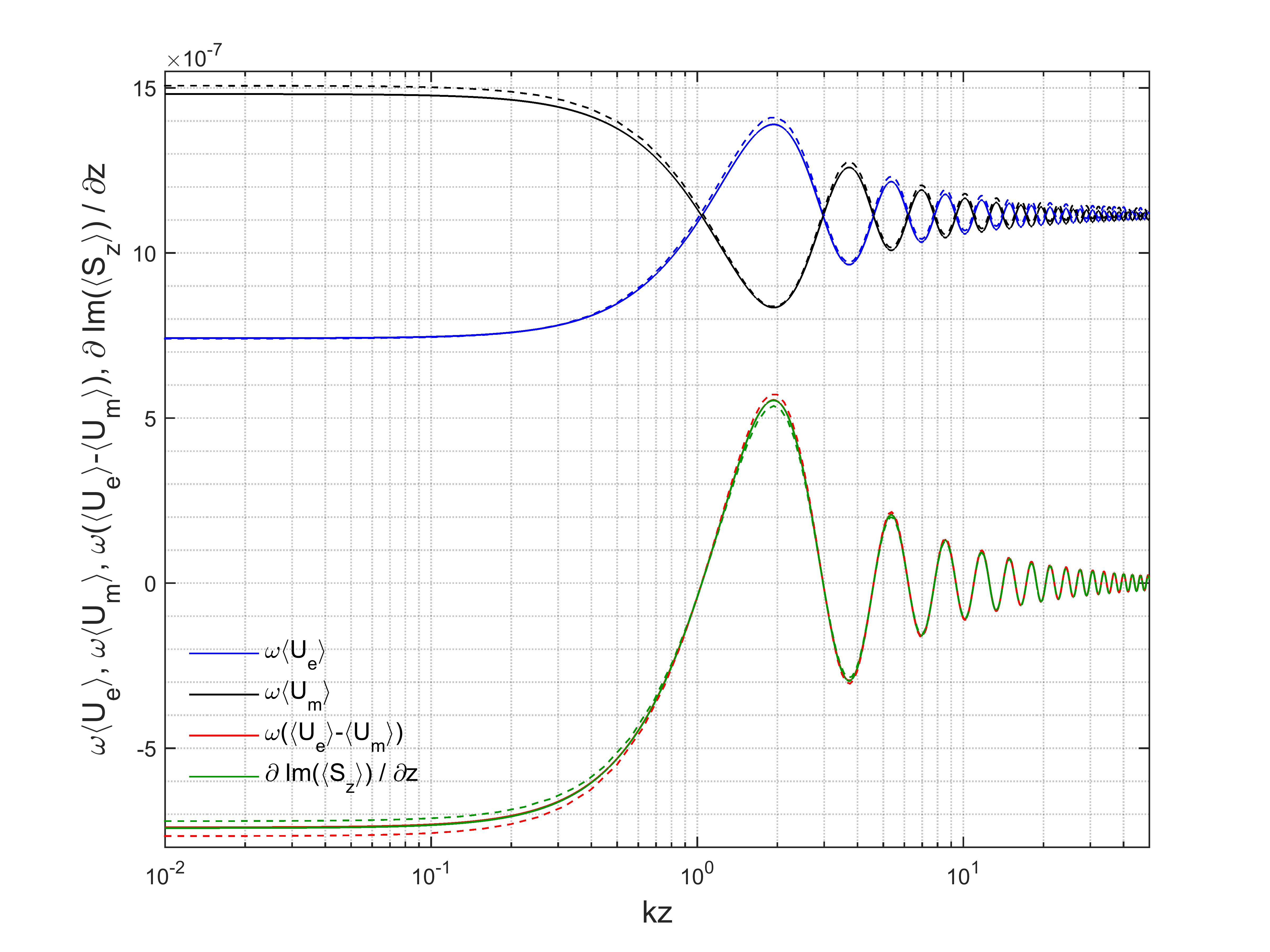}
\end{center}
{
\caption{\label{fig:conservenergy_avgSz} \small
Conservation of mean linear power flux and rate of change of energy density $\ul{\nabla} \cdot \langle \ul{S} \rangle \equiv \rmj \partial {\rmIm} (\langle{S_z}\rangle) / \partial z= \rmj \omega (\langle U_e \rangle - \langle  U_m \rangle)$ [in units W/m$^3$] as a function of height $kz$ above a PEC boundary for $\langle{{\cal E}^\prime_0}^2 \rangle = \langle{{\cal E}^{\prime\prime}_0}^2 \rangle = 10^{-4}$ (V/m)$^2$ at $f=100$ MHz. Solid: MC simulation; dashed: theory.
}
}
\end{figure}

For deterministic fields in a lossless medium, the Poynting theorem states that
\bea
\ul{\nabla} \cdot \ul{S} + \rmj \omega ( U_m - U_e) = 0. \label{eq:conserv_S_determ}
\eea
Application of (\ref{eq:EalphaHbeta}) yields
\bea
\frac{\partial}{\partial z} 
\left \{
\begin{array}{l}
E_x H^*_y\\
E_y H^*_x
\end{array}
\right \}
&=& 
\pm \frac{\rmj 2 k}{3} \delta(\Omega_1\Delta\Omega_2) 
\left (
\cE_{1\theta} \cH^*_{2\phi} - \cE_{1\phi} \cH^*_{2\theta} \right )
\nonumber\\
&~&\times
\left [ j_0(2kz) -2 j_2(2kz) \right ]
\label{eq:dSzdz}
\eea 
where the upper and lower signs refer to $E_x H^*_y$ and $E_y H^*_x$, respectively. Substituting (\ref{eq:EthetaEphi})--(\ref{eq:HthetaHphi}) herein
followed by ensemble averaging yields $\partial \langle S_z \rangle / \partial z $, which is found to coincide with the result obtained by direct differentiation of (\ref{eq:finalavgS}). Thus, $\langle \ul{\nabla} \cdot \ul{S} \rangle = \ul{\nabla} \cdot \langle \ul{S} \rangle$. 
Moreover, by differentiation of (\ref{eq:avgUeUm}) using 
\bea
\frac{{\rmd}j_n(2kz)}{{\rmd}z} 
= \frac{2k}{2n+1} \left [ n j_{n-1}(2kz) - (n+1) j_{n+1} (2kz) \right ]
\label{eq:recurr_diff_Besselj}
\eea
it is found that (\ref{eq:conserv_S_determ}) also applies to $\langle S \rangle$, viz.,
\bea
\ul{\nabla} \cdot \langle \ul{S} \rangle 
&=& - \rmj \omega \frac{2\eps_0 \langle |{\cal E}_0|^2 \rangle}{3} \left [ j_0(2kz) - 2 j_2(2kz) \right ]\nonumber\\ 
&=& - \rmj \omega (\langle U_m \rangle - \langle U_e \rangle).
\label{eq:conserv_avgS}
\eea 
The result (\ref{eq:conserv_avgS})
is confirmed by MC simulation shown in Fig. \ref{fig:conservenergy_avgSz}. 
Thus, Poynting's theorem for (physical) deterministic power and energy extends to their (arithmetic) averages for random fields, and also to the average power flux (divergence).  
The extended theorem indicates that, for arbitrary $kz$, the spatial local flow of the reactive (nonradiating) linear and angular average power fluxes $\rmc^2 \langle {P} \rangle$ and $\rmc^2 \langle {M} \rangle$ are associated and in phase with temporal oscillations of the imbalance between electric and magnetic average energies.
For increasing $kz$, the sign of their difference (i.e., the dominance of local average magnetic over electric energy density, or vice versa) alternates, while the magnitude (i.e., strength of the imbalance) decreases. 

%**************FIGURE 1006****************
%**************FIGURE ****************

\subsection{Conservation of Average LM and AM\label{sec:MST}}
\subsubsection{Average EM Stress}
\begin{figure}[htb] \begin{center} \begin{tabular}{c}
\vspace{-0.8cm}\\
\hspace{-0.75cm}
\includegraphics[scale=0.50]{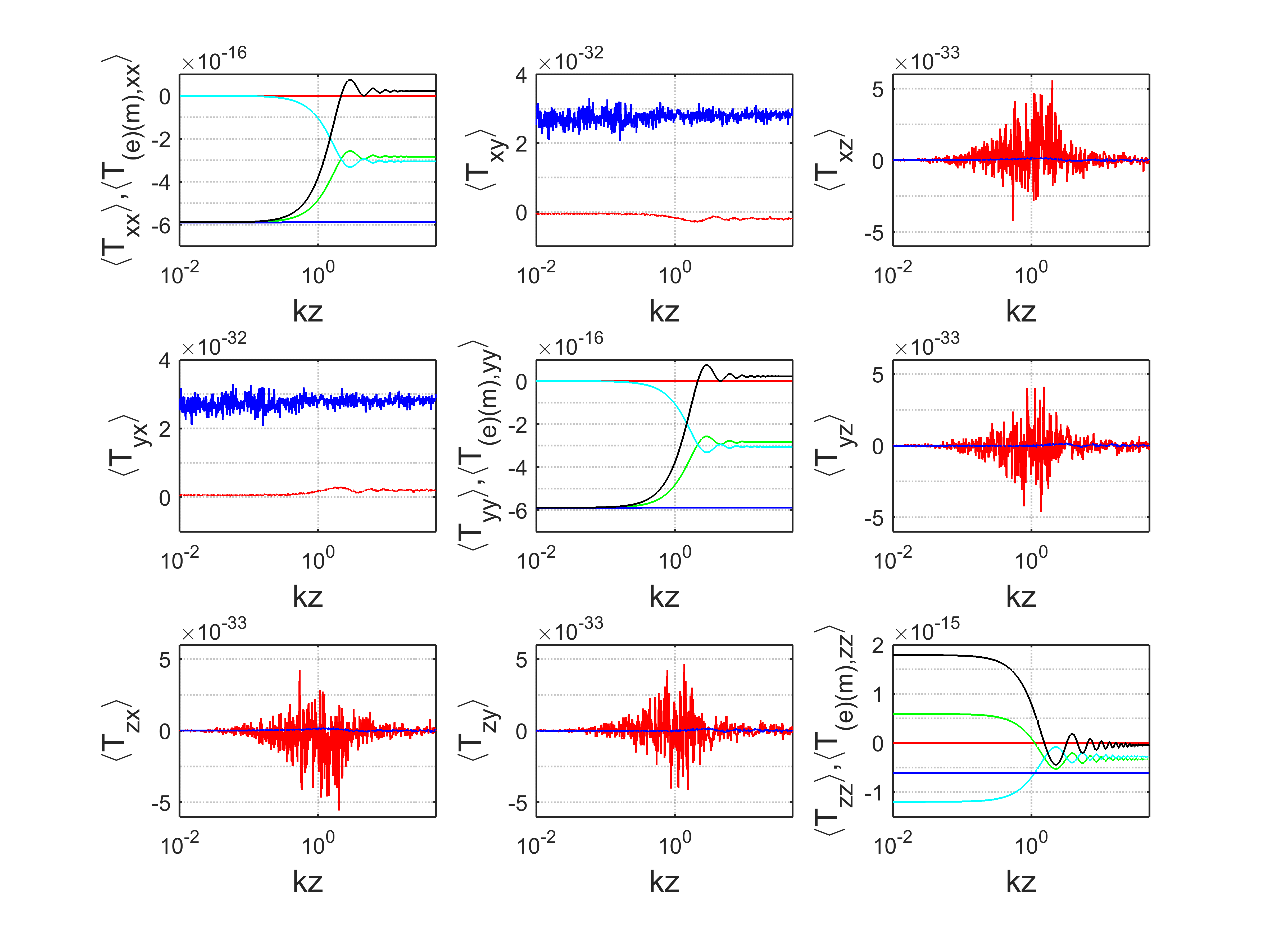}\\
\vspace{-1cm}\\
\end{tabular}
\end{center}
{
\caption{\label{fig:avgT} \small
MC simulated $\langle\dul{T}\rangle$ and $\langle\dul{T}_{(e)(m)}\rangle$ [in units J/m$^3$ or N/m$^2$] as a function of height $kz$ above a PEC boundary for $\langle{{\cal E}^\prime_0}^2 \rangle = \langle{{\cal E}^{\prime\prime}_0}^2 \rangle = 10^{-4}$ (V/m)$^2$.
Blue: Re[$\langle T_{\alpha\beta}\rangle$];
red: Im[$\langle T_{\alpha\beta}\rangle$];
green: Re[$\langle T_{e,\alpha\alpha}\rangle$];
cyan: Re[$\langle T_{m,\alpha\alpha}\rangle$];
black: Re[$\langle T_{e,\alpha\alpha}\rangle-\langle T_{m,\alpha\alpha}\rangle$].
\\
}
}
\end{figure}

For stochastic fields, the (random) symmetrized Maxwell stress dyadic (cf. Appendix)
\bea
\dul{T}(\ul{r},k) &=& \dul{T}_e + \dul{T}_m \stackrel{\Delta}{=} \frac{\eps_0}{2} \left ( \ul{E} \, \ul{E}^* - \frac{1}{2} |\ul{E}|^2 \right ) \nonumber\\
&~&
{+} \frac{\mu_0}{2} \left ( \ul{H}^* \ul{H} 
 - \frac{1}{2} |\ul{H}|^2 \dul{I} 
 \right ) 
\label{eq:T_cw}
\eea
characterizes the random radiation pressure $\dul{T} \cdot \ul{1}_z$ (LM flux) and shear $\dul{T} \times \ul{1}_z$ with reference to the surface normal $\ul{1}_z$. 
The resultant exerted random EM force follows by integrating $\dul{T}$ across the oriented boundary surface (or $\ul{\nabla} \cdot \dul{T}$ across the enclosed volume).
To evaluate $\langle \dul{T} \rangle$, note that 
\bea
\langle E_\alpha E^*_\beta \rangle = \langle H_\alpha H^*_\beta \rangle = 0,~~~{\forall}\alpha \not = \beta \in \{x,y,z\}
\label{eq:avgEalphaEbetap}
\eea
because the azimuthal dependence of their kernels are of the form $\sin \phi$, $\cos\phi$ or $\sin(2\phi)$, or because $\langle {\cal E}_{1\phi}{\cal E}^*_{2\theta}\rangle \propto \langle \sin \psi_1 \cos\psi_2 \rangle = 0$. 
From (\ref{eq:T_cw}) and (\ref{eq:avgEalphaEbetap}), with the aid of \cite[eqs. (8)--(10)]{arnaTEMCmay2006}, it follows that $\langle \dul{T}\rangle$ is diagonal, isotropic, and homogeneous (i.e., nondispersive), viz., 
\bea
\langle \dul{T}(kz) \rangle 
&=& -\frac{\langle U_{em,z} \rangle}{2} \dul{I}_t + \frac{\langle U_{em,z} \rangle - \langle U_{em,t} \rangle}{2} \ul{1}_z \ul{1}_z \nonumber\\ 
&=& - \frac{\eps_0 \langle |{\cal E}_0|^2 \rangle}{3} \dul{I} 
\label{eq:avg_dyadicT}
\eea
where $\dul{I}_t \stackrel{\Delta}{=} \ul{1}_x \ul{1}_x + \ul{1}_y \ul{1}_y$ is the transverse unit dyadic, with
\bea
\langle U_{em,z} \rangle &\stackrel{\Delta}{=}& \langle U_{e,z} \rangle + \langle U_{m,z} \rangle = \frac{2}{3} \langle \eps_0 |{\cal E}_0|^2 \rangle, \nonumber\\
\langle U_{em,t} \rangle &\stackrel{\Delta}{=}& \langle U_{e,t} \rangle + \langle U_{m,t} \rangle = \frac{4}{3} \langle \eps_0 |{\cal E}_0|^2 \rangle,~\forall kz \geq 0~~~~
\eea
and $\langle U_{(e)(m),t} \rangle \stackrel{\Delta}{=} \langle U_{(e)(m),x} \rangle + \langle U_{(e)(m),y} \rangle$.
The negative sign in (\ref{eq:avg_dyadicT}) indicates that the average stress constitutes a pressure, as opposed to a tension.
While all $\langle U_{(e)(m),(\alpha)(t)} \rangle$ are individually dispersive, the sum of each matching pair $\langle U_{e,\alpha} \rangle + \langle U_{m,\alpha} \rangle $ -- and hence $\langle U_{em,t} \rangle$, $\langle U_{em,z} \rangle$ and $\langle \dul{T}\rangle$  -- are real and dispersionless with respect to $kz$. Thus, 
\bea
\langle \dul{T} \rangle = \langle T \rangle \dul{I} = \left \langle \dul{\overline{\rmT}} \right \rangle 
\label{eq:Tidentities1}
\eea
and
\bea
\ul{\nabla} \cdot \langle \dul{T} \rangle = \ul{0},~~~\frac{\partial \langle {T}_{\alpha\alpha} \rangle}{\partial z} = {0},~~~\forall \alpha\in\{x,y,z\}.
\label{eq:Tidentities2}
\eea
Fig. \ref{fig:avgT}
shows MC results for $\langle \dul{T}(kz)\rangle$. 
The diagonal elements are easily verified to correspond to their theoretical constant real values (\ref{eq:avg_dyadicT}).
The residual off-diagonal $\langle T_{\alpha\beta}\rangle$ originate from finite-precision errors in the numerical quadrature of (\ref{eq:avgEalphaEbetap}), yet demonstrate that $\langle \dul{T} \rangle$ is Hermitean.

Unlike the overall $\langle \dul{T} \rangle$, the average individual electric $\langle \dul{T}_e\rangle$ and magnetic $\langle \dul{T}_m\rangle$ stress dyadics, i.e.,
\bea
\langle \dul{T}_{(e)(m)}(kz) \rangle 
&=& -\frac{\langle U_{(e)(m),z} \rangle}{2} \dul{I}_t \nonumber\\
&~& + \frac{\langle U_{(e)(m),z} \rangle - \langle U_{(e)(m),t} \rangle}{2} \ul{1}_z \ul{1}_z ~~~
\label{eq:avg_dyadicTe}
\label{eq:avg_dyadicTm}
\eea
are dispersive, as follows from (\ref{eq:avgUexyUmxy})--(\ref{eq:avgUezUmz}).
These are also shown in Fig. \ref{fig:avgT}.
From Fig. \ref{fig:Uenv_all} and (\ref{eq:avg_dyadicTm}), it follows that the average normal electric stress $\langle T_{e,zz} \rangle$ changes from tension ($>0$) to pressure ($<0$) at $kz\simeq 1.14$. By contrast, the average normal magnetic stress always occurs as a pressure ($\langle T_{m,zz} \rangle < 0$). 
%**************FIGURE 61****************
%**************FIGURE ****************

As a practical EMMC application of $\dul{T}$ within the realm of mode-stirred reverberation, consider the concept of electromechanical {\it self-stirring}. In this scenario, LM or AM arises from the EM stress caused by a source field impinging onto suspended or free-flowing small scatterers, causing their motion or morphing if inertia is sufficiently small. In turn, this affects the cavity field distribution, hence $\dul{T}(\ul{r})$ via (\ref{eq:T_cw}), and therefore $\ul{P}(\ul{r})$, $\ul{M}(\ul{r},\ul{r}_0)$ via (\ref{eq:conserv_P_determ})--(\ref{eq:conserv_M_determ}), etc. 
The result is akin to that of chaff in radar, except that the mechanism for its dynamics here is purely EM and equally feasible in vacuum.
The efficiency and control of self-stirring is governed by the field strength and  mechanical properties of each scatterer. 
Contrary to conventional mechanical mode stirring, self-stirring is most efficient for electrically {\em small\/} scatterers, as it relies on a net nonzero integrated $\dul{T}$ across their surface.
As is well known, mechanical pressure exerted onto the walls of an overmoded microwave cavity easily results in substantial changes to mode degeneracy, coupling and spectra, even for geometric distortions smaller than $\lambda/100$ \cite[Fig. 1]{schroe1954}. A fortiori, corresponding effects of EM stress are relevant to macro- or mesoscale structures with small inertia. 
If left uncontrolled, self-stirring produces noise additional to thermal noise caused by ohmic dissipation.

\subsubsection{Conservation of LM and AM}
For deterministic fields in vacuum, the conservation of LM density states that \cite{jack1975}, \cite{kong1990}
\bea
\ul{\nabla} \cdot \dul{T}  - \rmj \omega \ul{P} = \ul{0}
\label{eq:conserv_P_determ}
.
\eea
This follows from $\ul{P}$ in (\ref{eq:Mcomplex}) upon adding $(\ul{\nabla} \cdot \ul{H})\ul{H}^*\equiv \ul{0}$, dual-symmetrizing $\eps_0\ul{\nabla}|\ul{E}|^2$, and recalling that $\ul{\nabla}|\ul{E}|^2= \ul{\nabla} \cdot (|\ul{E}|^2\dul{I})$ and $(\ul{\nabla}\cdot \ul{E})\ul{E}^* = \ul{\nabla}\cdot(\ul{E}\,\ul{E}^*)$.
The corresponding conservation of AM follows by pre-multiplying (\ref{eq:conserv_P_determ}) by $\ul{r} \times$ and using the dyadic identities
$
\ul{r} \times (\ul{\nabla} \cdot \dul{T}) = 
-\ul{r} \cdot (\ul{\nabla} \times \dul{T}) =
-\ul{\nabla} \cdot \left ( \dul{T} \times \ul{r} \right )
$
to yield
\bea
\ul{\nabla} \cdot \left ( \dul{T} \times \ul{r} \right ) + \rmj \omega \ul{M} = \ul{0}.
\label{eq:conserv_M}
\label{eq:conserv_M_determ}
\eea
For arbitrary $kz$ from the PEC boundary, (\ref{eq:conserv_M_determ}) is in fact a manifestation of Noether's theorem with generator $\ul{1}_z \times \ul{r} \equiv r \ul{1}_\phi$, which holds because of rotational symmetry around $\ul{1}_z$, such that $M_z(kz)=0$ -- as already found in (\ref{eq:finalavgM}) -- and hence $[ \ul{\nabla} \cdot \left ( \dul{T}(kz) \times \ul{r} \right ) ]_z = 0$ follows from (\ref{eq:conserv_M_determ}).

Since $\ul{P}$ and $\ul{M}$ are proportional to linear and angular EM power flux, eqs. (\ref{eq:conserv_P_determ}) and (\ref{eq:conserv_M_determ}) provide a conduit between EM and mechanical effects. 
On account of Gauss's theorem, the rate of change of the LM inside a finite volume can thus be observed from the net flux of stress $\ul{\nabla} \cdot \dul{T}$ through its boundary where EM forces act. 
For random fields, it appears that the average flux $\langle \ul{\nabla} \cdot \dul{T} \rangle$ and the fluxes of the average $\ul{\nabla} \cdot \langle \dul{T}_{(e)(m)} \rangle$ enable alternative physical interpretations of $\langle \ul{P} \rangle$ in (\ref{eq:finalavgS}), as will be shown next.

\paragraph{Average Flux}
To extend and evaluate (\ref{eq:conserv_P_determ})--(\ref{eq:conserv_M_determ}) for random fields near a PEC plane, note that spatial variations of $E_\alpha E^*_\beta$ and $H^*_\alpha H_\beta$ are in the normal direction 
($\ul{\nabla}=\ul{1}_z(\partial/\partial z)$) for all $\alpha,\beta \in \{x,y,z\}$, whence %for (\ref{eq:conserv_P})
\bea
\ul{\nabla} \cdot \dul{T} = \sum_{\alpha=z} \sum_{\beta=x,y,z} \frac{\partial{T}_{\alpha\beta}}{\partial \alpha} \ul{1}_\beta 
.
~~~
\label{eq:divT}
\eea
Correspondingly, the dyadic skew product $\dul{T} \times\ul{r}$ may be calculated termwise, i.e., $(\ul{E} \,\ul{E}^*) \times \ul{r} = \ul{E} (\ul{E}^* \times \ul{r})$, etc., followed by differentiation, resulting in 
\bea
&~&
\ul{r} \times ( \ul{\nabla} \cdot \dul{T} ) 
=
\sum_{\alpha=z} \sum_{\beta=x,y,z} (\ul{r} \times \ul{1}_\beta ) \frac{\partial{T}_{\alpha\beta}}{\partial \alpha} 
.
\label{eq:rcrossdivT}
~~~~~~
\eea
It can be easily shown from (\ref{eq:EalphaHbeta}) that 
$\langle T_{zx} \rangle = \langle T_{zy} \rangle =0=\langle T_{(e)(m),zx}\rangle =\langle T_{(e)(m),zy} \rangle$ and that
\bea
&~&\hspace{-0.5cm}
\left \langle E_\alpha \frac{\partial E^*_\beta}{\partial z} \right \rangle = 
\left \langle H_\alpha \frac{\partial H^*_\beta}{\partial z} \right \rangle = 0,~
{\forall}\alpha \not = \beta \in \{x,y,z\}
~~~~~~
\label{eq:avgddiffEalphaEbetap}
\eea
for the same reasons as those given for (\ref{eq:avgEalphaEbetap}).
Therefore, $\langle {\partial T_{zx}}/{\partial z} \rangle = \langle {\partial T_{zy}}/{\partial z}\rangle = 0$.
Ensemble averaging of (\ref{eq:divT}) and (\ref{eq:rcrossdivT}) yields 
\bea
\langle \ul{\nabla} \cdot \dul{T} \rangle &=& \langle \left ( \ul{\nabla} \cdot \dul{T} \right )_z \rangle = 
\left \langle \frac{\partial T_{zz}}{\partial z} \right \rangle
\ul{1}_z
\label{eq:avgdivT}\\
\langle \ul{r} \times ( \ul{\nabla} \cdot \dul{T} ) \rangle &=& 
y \left \langle \frac{\partial T_{zz}}{\partial z} \right  \rangle  
\ul{1}_x - x \left \langle \frac{\partial T_{zz}}{\partial z} \right \rangle
\ul{1}_y . \label{eq:avgrtimesdivT}
\eea
Thus, whereas $\ul{\nabla}\cdot {\dul{T}}$ and $\ul{r} \times (\ul{\nabla}\cdot {\dul{T}})$ depend on the $z$-derivative of both the fluctuating EM stress ($T_{zz}$) and shear $(T_{zx}, T_{zy}$) components as generated by each plane wave component, only the dependence on the EM stress survives after averaging.

\begin{figure}[htb] \begin{center} \begin{tabular}{c}
\vspace{-0.5cm}\\
\hspace{-0.75cm}
\includegraphics[scale=0.50]{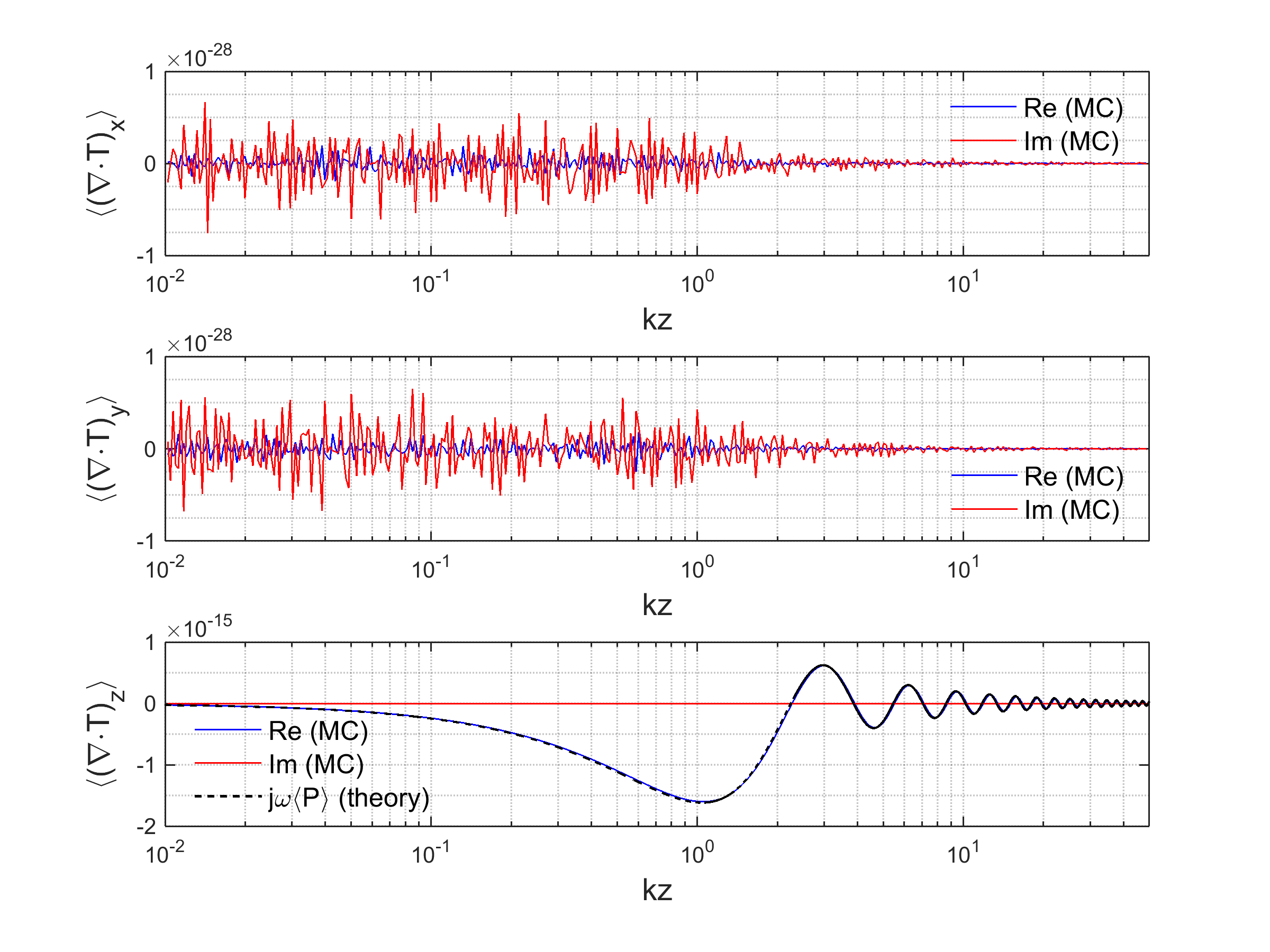}\\
\vspace{-0.5cm}\\
\end{tabular}
\end{center}
{
\caption{\label{fig:avgdivT_z} \small 
Shear ($x,y$) and normal ($z$) components of average spatial gradient $\langle \ul{\nabla} \cdot \dul{T} \rangle$ [in units J/m$^4$ or N/m$^3$] as a function of height $kz$ above a PEC boundary for $\langle{{\cal E}^\prime_0}^2 \rangle = \langle{{\cal E}^{\prime\prime}_0}^2 \rangle = 10^{-4}$ (V/m)$^2$ at $f=100$ MHz. 
Blue solid: MC real part; 
red solid: MC imaginary part; 
blue dashed, theoretical $\rmj\omega \langle {P}_z \rangle$.
}
}
\end{figure}

\begin{figure}[htb] \begin{center} \begin{tabular}{c}
%\vspace{-0.5cm}\\
\hspace{-0.75cm}
\includegraphics[scale=0.50]{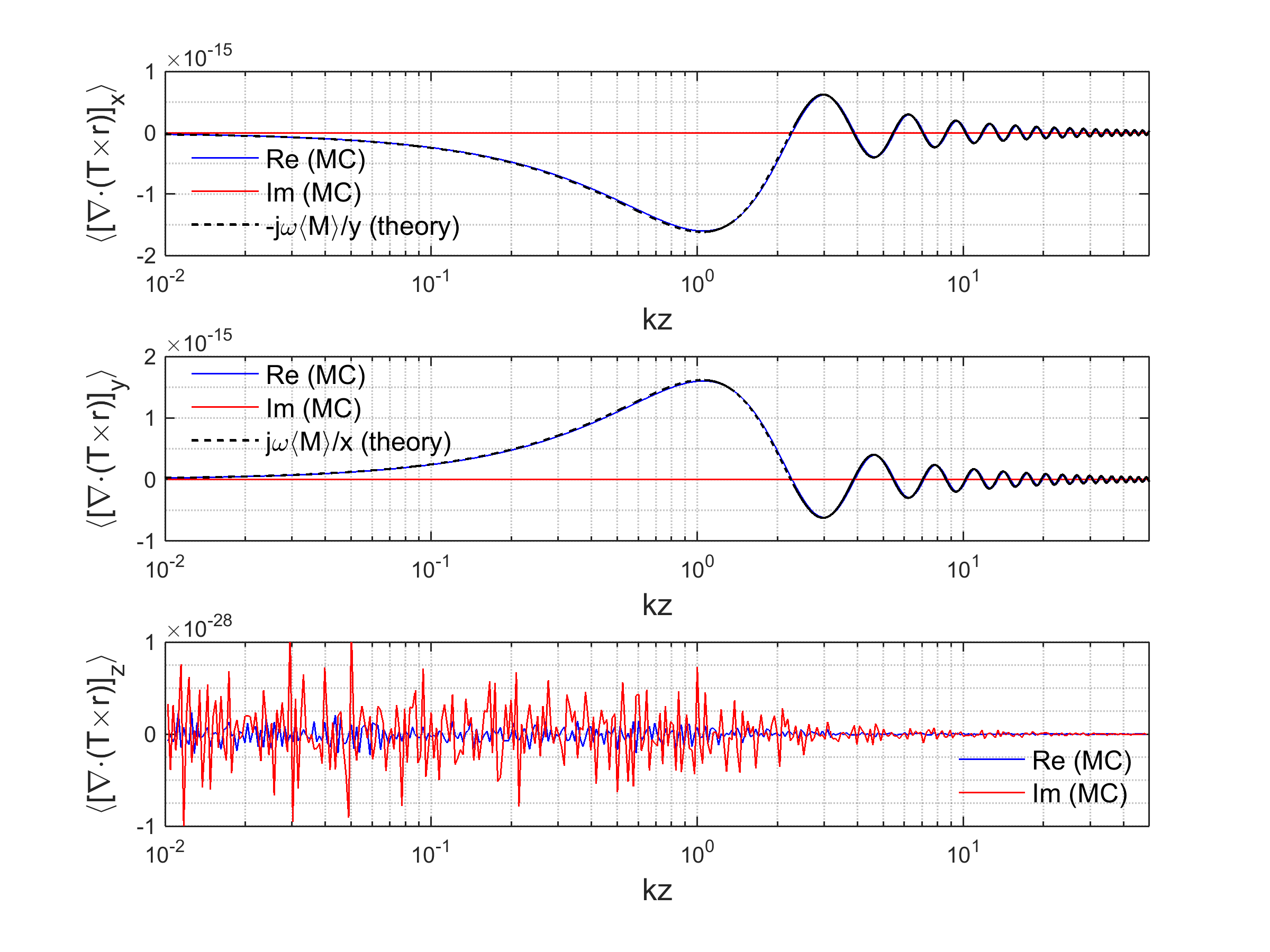}
\vspace{-0.5cm}\\
\end{tabular}
\end{center}
{
\caption{\label{fig:avgdivTcrossR} \small
Angular shear ($x,y$) and stress ($z$) components of average spatial gradient $\langle \ul{\nabla} \cdot ( \dul{T} \times \ul{r} ) \rangle$ [in units J/m$^3$ or N/m$^2$] as a function of height $kz$ above a PEC boundary for $\langle{{\cal E}^\prime_0}^2 \rangle = \langle{{\cal E}^{\prime\prime}_0}^2 \rangle = 10^{-4}$ (V/m)$^2$ at $f=100$ MHz.
Blue solid: MC real part;
red solid: MC imaginary part;
blue dashed: theoretical $\rmj \omega \langle {M}_{(x)(y)} \rangle$.
}
}
\end{figure}

Figs. \ref{fig:avgdivT_z} and \ref{fig:avgdivTcrossR} show $\langle \ul{\nabla} \cdot \dul{T} \rangle$ and $\langle \ul{\nabla} \cdot ( \dul{T} \times  \ul{r} ) \rangle$ as a function of $kz$, respectively, for a MC simulation based on $\dul{T}$ calculated from (\ref{eq:EalphaHbeta}) and after finite differencing of $\nabla_z$ for discrete $kz$.
It is seen that conservation of LM and AM also holds between their statistical averages and the average flux of EM stresses and their moments (or torque, in case of SAM), respectively, i.e., 
\bea
\langle \ul{\nabla} \cdot \dul{T} \rangle - \rmj \omega \langle \ul{P} \rangle &=& \ul{0}
\label{eq:conserv_avgP}\\
\langle\ul{\nabla} \cdot \left ( \dul{T} \times \ul{r} \right ) \rangle + \rmj \omega \langle \ul{M} \rangle &=& \ul{0}.
\label{eq:conserv_avgM}
\eea
While (\ref{eq:conserv_avgP}) and (\ref{eq:conserv_avgM}) follow of course trivially from (\ref{eq:conserv_P_determ}) and (\ref{eq:conserv_M_determ}) from a purely mathematical perspective, the demonstration of their validity here through an independent MC calculation of $\dul{T}$, $\langle \dul{T} \rangle$, $\langle \ul{\nabla} \cdot \dul{T} \rangle$ and $\langle \ul{\nabla} \cdot (\dul{T} \times \ul{r} )\rangle$ starting from the plane-wave expansion serves to validate (\ref{eq:conserv_P_determ}) and (\ref{eq:conserv_M_determ}) as a starting point based on first principles.
 
%**************FIGURE 62****************
%**************FIGURE ****************

%**************FIGURE 63****************
%**************FIGURE ****************

\paragraph{Flux of Average Electric and Magnetic Stresses}
From (\ref{eq:finalavgS}), (\ref{eq:Tidentities2})
and (\ref{eq:conserv_P_determ}), it follows that in general
\bea
\ul{\nabla} \cdot \langle \dul{T} \rangle - \rmj \omega \langle \ul{P} \rangle &\not =& \ul{0}
\label{eq:NOTconserv_avgP}\\
\ul{\nabla} \cdot \langle \dul{T} \times \ul{r} \rangle + \rmj \omega \langle \ul{M} \rangle &\not =& \ul{0}
\label{eq:NOTconserv_avgM}
\eea
except at the optimum locations $k z_s$ where $\langle \ul{P} \rangle$ or $\langle \ul{M} \rangle$ vanish (cf. Sec. \ref{sec:optimum}).
Generally, spatial differentiation and statistical averaging operations commute (in particular $\langle\ul{\nabla} \cdot \dul{T} \rangle = \ul{\nabla} \cdot \langle \dul{T} \rangle$) provided that the probability distribution (of $\dul{T}$) is spatially homogeneous \cite[Sec. 7.2]{groo1969}. This condition has previously been shown to be violated for $U_e$ near an EM boundary in its normal direction \cite[Figs. 5-7]{arnaRS}. Hence 
(\ref{eq:NOTconserv_avgP}) and (\ref{eq:NOTconserv_avgM}) are consistent with this result.
On the other hand, in Sec. \ref{sec:Poynting} it was found that $\langle \ul{\nabla} \cdot \ul{S} \rangle = \ul{\nabla} \cdot \langle \ul{S} \rangle$ even though $S$ and $\langle S \rangle$ (as well as $U_m - U_e$ and $\langle  U_m \rangle - \langle U_e \rangle$) are dispersive near the boundary.
This result is again not inconsistent with commutativity, because the boundary zone field is statistically inhomogeneous. Note, however, that the average total energy density is nondispersive, viz., $\langle U_{em} \rangle \equiv \langle U_e\rangle + \langle U_m\rangle = 2 \eps_0 \langle |{\cal E}_0|^2 \rangle$.

\begin{figure}[htb] \begin{center} \begin{tabular}{c}
\vspace{-0.5cm}\\
\hspace{-0.75cm}
\includegraphics[scale=0.50]{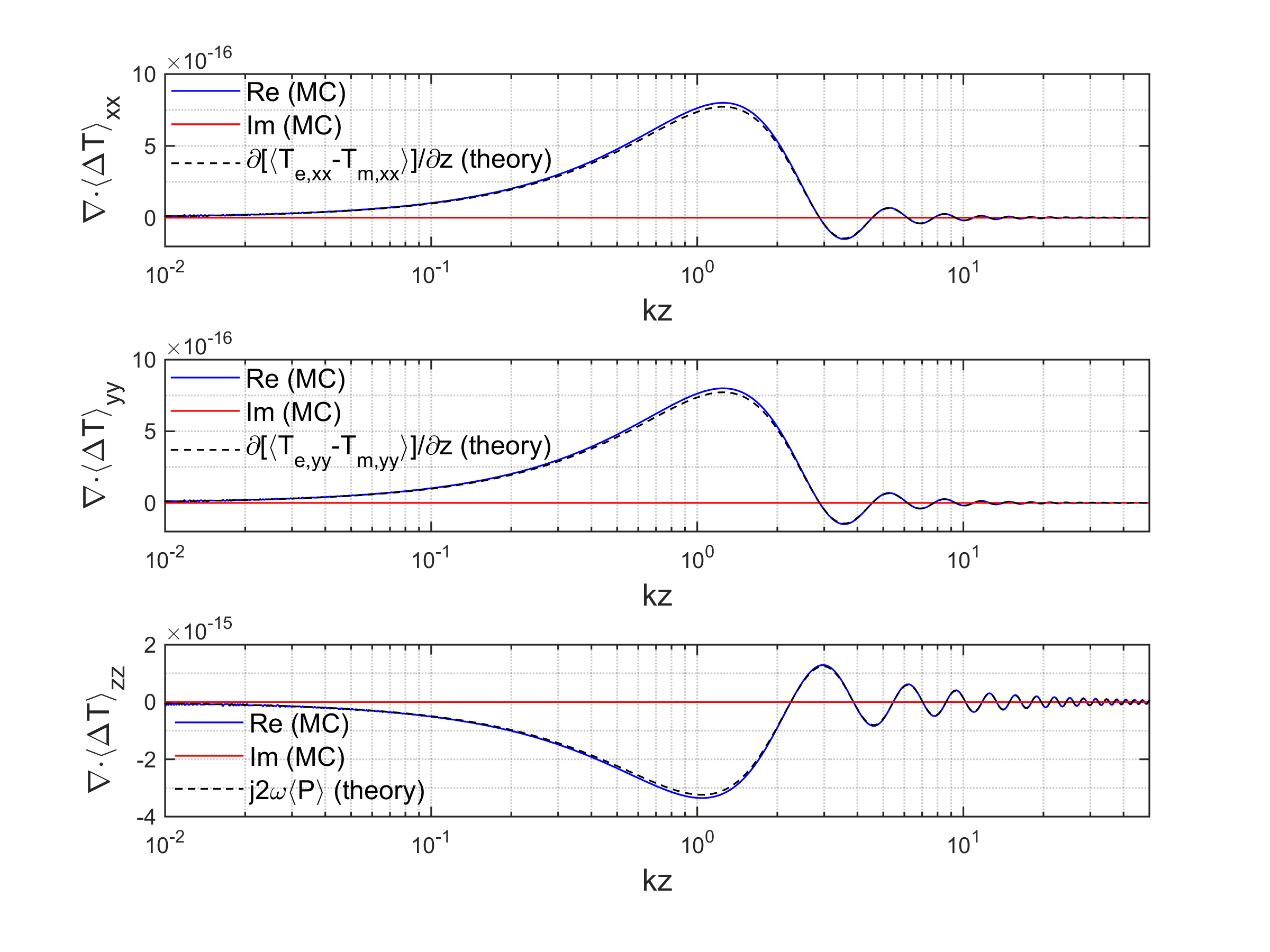}\\
\vspace{-0.5cm}\\
\end{tabular}
\end{center}
{
\caption{\label{fig:diffavgTdiff} \small
Spatial gradients of average differential EM stress components $\ul{\nabla} \cdot \langle \Delta \dul{T} \rangle$ [in units J/m$^4$ or N/m$^3$] as a function of height $kz$ above a PEC boundary and $\langle{{\cal E}^\prime_0}^2 \rangle = \langle{{\cal E}^{\prime\prime}_0}^2 \rangle = 10^{-4}$ (V/m)$^2$ at $f=100$ MHz.
Blue solid: MC real part, red solid: MC imaginary part, black dashed: theoretical $\partial \langle {T}_{e,\alpha\alpha} {-} {T}_{m,\alpha\alpha} \rangle / \partial z$ for $\alpha = x,y,z$ [eq. (\ref{eq:davgTemdz_final})].
}
}
\end{figure}

Expressions for $\ul{\nabla} \cdot \langle \dul{T}_{(e)(m)} \rangle$ and $\ul{\nabla} \cdot \langle \dul{T}_{(e)(m)} \times \ul{r} \rangle$ are readily obtained as follows.
Application of (\ref{eq:EalphaHbeta}) to the calculation of 
$\partial E_\alpha E^*_\beta/\partial{z}$ and $\partial H_\alpha H^*_\beta/\partial{z}$ produces zero when $\alpha \not = \beta$, and
\bea
&~& \hspace{-1cm}
\frac{\partial}{\partial z} 
\left \{ 
\begin{array}{l}
{|E_{(x)(y)}|^2}\\
{|H_{(x)(y)}|^2}
\end{array}
\right \}  
=
\pm \frac{2 k}{5} \delta(\Omega_1\Delta\Omega_2) 
\left [
\left \{
\begin{array}{l}
\cE_{1\theta} \cE^*_{2\theta} \\
\cH_{1\theta} \cH^*_{2\theta}
\end{array}
\right \}
\right. \nonumber\\&~&\hspace{-1cm} \left. \times
\left [ 3 j_1(2kz) - 2 j_3(2kz) \right ]
+
\left \{
\begin{array}{l}
{\cE_{1\phi} \cE^*_{2\phi}} \\
{\cH_{1\phi} \cH^*_{2\phi}}
\end{array}
\right \}
5 j_1(2kz)
\right ]
\label{eq:dUexydz}
\label{eq:dUmxydz}
\eea 
\bea
%&~&
\frac{\partial}{\partial z} 
\left \{
\begin{array}{l}
{|E_z|^2}\\
{|H_z|^2}
\end{array}
\right \}
&=& 
\mp \frac{8k}{5} \delta(\Omega_1\Delta\Omega_2) 
\left \{
\begin{array}{l}
{\cE_{1\theta} \cE^*_{2\theta}} \\
{\cH_{1\theta} \cH^*_{2\theta}}
\end{array}
\right \}
\nonumber\\
&~&\times
\left [ j_1(2kz) + j_3(2kz) \right ]
\label{eq:dUezdz}
\label{eq:dUmzdz}
\eea 
otherwise. 
Substituting (\ref{eq:EthetaEphi})--(\ref{eq:HthetaHphi}) followed by ensemble averaging leads to
\bea
\frac{\partial \langle U_{(x)(y)} \rangle}{\partial z}
&=& \pm \frac{k \eps_0 \langle | {\cal E}_0 |^2 \rangle}{5} 
\left [ 4 j_1(2kz) - j_3(2kz)\right ]
\label{eq:dUxydz_final}
\\
\frac{\partial \langle U_{z} \rangle}{\partial z}
&=& \mp \frac{2k \eps_0 \langle |{\cal E}_0 |^2 \rangle}{5} 
\left [ j_1(2kz) + j_3(2kz)\right ]
\label{eq:dUzdz_final}\\
\frac{\partial \langle \dul{T}_{(e)(m)}\rangle}{\partial z} &=& 
\pm \frac{k \eps_0 \langle |{\cal E}_0 |^2 \rangle}{5}
\left \{ \left [ j_1(2kz) + j_3(2kz) \right ] \dul{I}_t 
\right. \nonumber\\ 
&~& \left. - 5 j_1(2kz) \ul{1}_z \ul{1}_z \right \}
\label{eq:davgTemdz_final}
\eea
where upper and lower signs again refer to electric ($U_e$, $\dul{T}_e$) and magnetic ($U_m$, $\dul{T}_m$) quantities, respectively. Eqs. (\ref{eq:dUxydz_final})--(\ref{eq:davgTemdz_final}) also follow from (\ref{eq:avgUexyUmxy})--(\ref{eq:avgUezUmz}) with the aid of (\ref{eq:recurr_diff_Besselj}).
Eq. (\ref{eq:davgTemdz_final}) confirms (\ref{eq:Tidentities2}) for the sum $\langle \dul{T} \rangle \equiv \langle \dul{T}_{e}\rangle + \langle \dul{T}_{m} \rangle$ and enables 
(\ref{eq:NOTconserv_avgP})--(\ref{eq:NOTconserv_avgM}) to be refined to
\bea
\ul{\nabla} \cdot \langle \dul{T}_{(e)(m)} \rangle 
&=& \mp \rmj \omega \langle \ul{P} \rangle 
= \mp k \eps_0 \langle |\cE_0|^2 \rangle j_1(2kz) \ul{1}_z 
\label{eq:divTem_final}
~~~~~\\
\ul{\nabla} \cdot \langle \dul{T}_{(e)(m)} \times \ul{r} \rangle 
&=& \pm \rmj \omega \langle \ul{M} \rangle
\label{eq:divcrossTem_final}
.
\eea
Furthermore, (\ref{eq:divTem_final}) and (\ref{eq:divcrossTem_final}) reveal a conservation law between the flux of the {\em differential\/} average EM stress $\Delta \langle \dul{T} \rangle = \langle \dul{T}_{e}\rangle - \langle \dul{T}_{m} \rangle$, and the rate of oscillation of the average LM or AM, viz.,
\bea
\ul{\nabla} \cdot ( \langle \dul{T}_{e} \rangle - \langle \dul{T}_{m} \rangle ) + \rmj 2 \omega \langle \ul{P} \rangle &=& \ul{0}\label{eq:conservP_avgTmminTe}\\
\ul{\nabla} \cdot \left [ ( \langle \dul{T}_{e} \rangle - \langle \dul{T}_{m} \rangle ) \times \ul{r} \right ] - \rmj 2 \omega \langle \ul{M} \rangle &=& \ul{0}.\label{eq:conservM_avgTmminTe}
\eea
Eq. (\ref{eq:conservP_avgTmminTe}) also follows more directly by substituting (\ref{eq:avgUetUmt}) and (\ref{eq:avgUezUmz}) into (\ref{eq:avg_dyadicTm}), which yields $\Delta{T}_{zz} = j_0(2kz)$, and applying (\ref{eq:recurr_diff_Besselj}).
To illustrate and validate these results, Figs. \ref{fig:avgT} and \ref{fig:diffavgTdiff}\footnote{The transverse diagonal elements ($\alpha=x,y$; cf. top two plots in Fig. \ref{fig:diffavgTdiff}) are only relevant in configurations for which ${\nabla}_x$ and/or ${\nabla}_y$ are also nonzero, e.g., when one or more additional adjacent boundaries are perpendicular to $\ul{1}_x$ or $\ul{1}_y$.
}
show $\langle \Delta T_{\alpha\alpha} (kz)\rangle \stackrel{\Delta}{=} \langle {T}_{e,\alpha\alpha} (kz) \rangle - \langle {T}_{m,\alpha\alpha} (kz) \rangle$ and their divergence, respectively. The latter Figure confirms (\ref{eq:conservP_avgTmminTe}) for $\alpha=z$ and (\ref{eq:davgTemdz_final}) for arbitrary $\alpha$.

%**************FIGURE 1062****************
%**************FIGURE ****************

\subsection{Relation Between EM Stress and Energy}
For deterministic fields, substitution of (\ref{eq:conserv_P_determ}) into (\ref{eq:conserv_S_determ}) enables the second-order spatial derivatives (curvatures) of the radiation stress components to be related to the EM energy imbalance $U_m-U_e$ (or the second-order time derivative of the total energy, for general time-dependent fields) without recourse to the LM or power, viz.,
\bea
\rmj \omega \ul{\nabla} \cdot \ul{P} = k^2 (U_m-U_e) = \ul{\nabla} \cdot (\ul{\nabla} \cdot \dul{T})
\eea
where $\ul{\nabla} \cdot (\ul{\nabla} \cdot \dul{T}) = \partial^2 T_{zz} / \partial z^2$ in the present configuration. 
For an averaged random field, the corresponding identity follows from (\ref{eq:conserv_avgS}) and (\ref{eq:conserv_avgP})  
as
\bea
\ul{\nabla} \cdot \langle \ul{\nabla} \cdot  \dul{T} \rangle - k^2 (\langle U_m \rangle - \langle U_e \rangle) = 0
.
\label{eq:secondorderPDEforT_gen}
\eea
Hence $\langle \ul{\nabla} \cdot \ul{\nabla} \cdot \dul{T} \rangle = \ul{\nabla} \cdot \langle \ul{\nabla} \cdot \dul{T} \rangle$ 
but $\ul{\nabla} \cdot \langle \ul{\nabla} \cdot \dul{T} \rangle \not = \ul{\nabla} \cdot \ul{\nabla} \cdot \langle \dul{T} \rangle = 0$
in view of the foregoing analysis, except asymptotically for $kz \rightarrow +\infty$ or at discrete locations and frequencies $kz_s$ at which the dispersion vanishes. 
In terms of the individual electric and magnetic average stresses and energies, (\ref{eq:secondorderPDEforT_gen}) with (\ref{eq:conservP_avgTmminTe}) degenerates into separate equations, viz.,
\bea
\ul{\nabla} \cdot \ul{\nabla} \cdot \langle \dul{T}_{(e)(m)} \rangle \pm k^2 \langle U_{(e)(m)} \rangle = 0\label{eq:secondorderPDEforTe_E}
\label{eq:secondorderPDEforTm_H}
\eea
where the upper and lower signs refer to $(\langle \dul{T}_e \rangle, \langle U_e \rangle)$ and $(\langle \dul{T}_m \rangle, \langle U_m \rangle)$, respectively.

\section{Conclusion}
The average LM (\ref{eq:finalavgS}) and AM (\ref{eq:finalavgM}) (and the corresponding linear and angular power flux densities) of a random field near a PEC boundary exhibit a dependence on the electric distance $kz$ that contains similarities as well as differences compared to the dependence of the average electric or magnetic energy densities (\ref{eq:avgUeUm})--(\ref{eq:avgUezUmz}). Similar to the energy,  
the strength of the average momentum and power  decays with increasing distance in a damped oscillatory manner and vanishes far away from the boundary. By contrast, the average momentum and power flux vanish on the boundary itself, their rate of decay with $kz$ is different, and their asymptotic deep values are attained at values of $kz$ in (\ref{eq:cond_S}) that are interleaved with (i.e., spaced by approximately $\pi/4$ from) those for the total energy density (\ref{eq:cond_U_e}). 
As an application to reverberation chambers, it was shown that by specifying a maximum tolerable deviation (nonuniformity) of the average boundary power or energy from its asymptotic free-space value, a performance based metric for the size of the working volume of a chamber can be  defined and calculated via (\ref{eq:WV_final}).

The Poynting theorem for conservation between energy imbalance and power flux in the absence of ohmic losses was found to remain satisfied for their (arithmetic) statistical averages in the case of random fields near a PEC plane, extending its validity beyond deterministic fields to (\ref{eq:conserv_avgS}).
Conservation of the average LM and AM can be expressed either for the average flux of the full EM stresses (\ref{eq:conserv_avgP}) and its moment (\ref{eq:conserv_avgM}), or in terms of the flux of electric or magnetic stress
(\ref{eq:divTem_final}) or their difference (\ref{eq:conservP_avgTmminTe}), and their moments (\ref{eq:divcrossTem_final}) and (\ref{eq:conservM_avgTmminTe}), respectively.
The average EM energy imbalance and stress are directly related as (\ref{eq:secondorderPDEforT_gen}) or individually as (\ref{eq:secondorderPDEforTe_E}).

As a final comment, LM and AM of random fields were found to exhibit a partial and statistical behaviour, rather than being an all-or-nothing property in the case of deterministic fields.
This is similar to other wave characteristics of inhomogeneous random fields, in particular for the degree of polarization \cite{arnaCE}, \cite{migl2016}.
This is not surprising in view of the fact that polarization content is already comprised as SAM within AM (\ref{eq:Ms_plus_Mo}).

\section*{Appendix: Maxwell Stress Dyadic for Time-Harmonic Random Fields}
The Maxwell stress dyadic is usually formulated in the time domain as (e.g., \cite[Sec. 6.8]{jack1975})
\bea
\dul{\rmT}(\ul{r},t) 
\stackrel{\Delta}{=} \eps_0 \ul{\rmE}\,\ul{\rmE} + \mu_0 \ul{\rmH}\,\ul{\rmH} - \frac{1}{2} \left ( \eps_0 \ul{\rmE}^2 + \mu_0 \ul{\rmH}^2 \right ) \dul{I} 
\label{eq:def_T}
.
\eea
With an assumed $\exp(\rmj \omega t)$ dependence, the corresponding expression for the complex $\dul{T}(\ul{r},\omega)$ for a time harmonic random field ($\ul{E},\ul{H},\omega)$ can be derived as follows. 
First, consider excitation by a single plane wave ($\ul{\cE},\ul{\cH}, \ul{k})$ from the angular spectrum of ($\ul{E},\ul{H})$. 
Being a sum of products of complex field quantities, the complex Lorentz force can be written as
\bea
\ul{\cF}(\ul{r},\ul{k}) = \frac{1}{2} \left ( \varrho \ul{\cE}^* + \ul{\cJ} \times \ul{\cB}^* \right ).
\label{eq:Lorentz}
\eea
Applying Gauss's and Amp\`{e}re's laws to express $\varrho$ and $\ul{\cJ}$ in terms of their source fields gives
\bea
\ul{\cF} = \frac{1}{2} \left [ (\ul{\nabla} \cdot \ul{\cD}) \ul{\cE}^* + (\ul{\nabla} \times \ul{\cH}) \times \ul{\cB}^* - \rmj \omega \ul{\cD} \times \ul{\cB}^* \right ].
\label{eq:F_intermed}
\eea
The last term in (\ref{eq:F_intermed}) equals $-\rmj \omega \ul{\cP}$. Using dyadic algebra and the vector identity $\ul{\cH}^* \times (\ul{\nabla} \times \ul{\cH}) = \ul{\nabla} ( \ul{\cH}^* \cdot \ul{\cH} ) - (\ul{\nabla} \cdot \ul{\cH}^* ) \ul{\cH} = \ul{\nabla}\, |\ul{\cH}|^2 - \ul{\nabla} \cdot ( \ul{\cH}^* \ul{\cH} )$, this yields in vacuum
\bea
\ul{\cF} &=& 
\frac{1}{2} \left [ \eps_0 \ul{\nabla} \cdot ( \ul{\cE} \, \ul{\cE}^* ) + 
\mu_0 \ul{\nabla} \cdot ( \ul{\cH}^* \ul{\cH} ) \right .\nonumber\\
&~& \left. - \mu_0 \ul{\nabla}  \cdot ( |\ul{\cH}|^2 \dul{I} ) \right ] - \rmj \omega \ul{\cP} =\ul{0}.
\label{eq:equilibrium}
\eea
Expressing $\rmj \omega \ul{\cP}$ as $\ul{\nabla} \cdot \dul{\cT}$ in view of (\ref{eq:conserv_P_determ}) gives
\bea
\dul{\cT}(\ul{r},\ul{k}) = \frac{1}{2} \left ( \eps_0 \ul{\cE} \, \ul{\cE}^* + \mu_0 \ul{\cH}^* \ul{\cH} - \mu_0 |\ul{\cH}|^2 \dul{I} \right ).
\label{eq:T_temp}
\eea
This expression for $\dul{\cT}$ is dual-asymmetric because (\ref{eq:Lorentz}) is as such.
Formally applying $\cH = \cE/\eta_0$ enables the last term in (\ref{eq:T_temp}) to be dual-symmetrized \cite[sec. 4.2]{lind1992} as 
$
\mu_0 |\ul{\cH}|^2 \dul{I} = \left ( \eps_0 |\ul{\cE}|^2 + \mu_0 |\ul{\cH}|^2\right ) \dul{I} / 2
$.
Spherical integration of each dyad $\cE_\alpha \cE^*_\beta$ and $\cH^*_\alpha \cH_\beta$ according to (\ref{eq:EalphaHbeta}) finally results in the stress dyadic for the time-harmonic random field as
\bea
\dul{T}(\ul{r},\omega) &=& \frac{1}{2} \left [ \eps_0 \ul{E} \, \ul{E}^* + \mu_0 \ul{H}^* \ul{H} \right. \nonumber\\
&~& \left. - \frac{1}{2} \left ( \eps_0 |\ul{E}|^2 + \mu_0 |\ul{H}|^2\right ) \dul{I} \right ]
.
\eea
\end{document}